# The Financial Bubble Experiment: Advanced Diagnostics and Forecasts of Bubble Terminations Volume I

D. Sornette, R. Woodard, M. Fedorovsky, S. Reimann, H. Woodard, W.-X. Zhou
(The Financial Crisis Observatory)*
*Department of Management, Technology and Economics,*
*ETH Zurich, Kreuzplatz 5, CH-8032 Zurich, Switzerland*

This is a summary of the first installment of the Financial Bubble Experiment (FBE), where we identified four asset bubbles in November and December 2009 and revealed their names on 3 May 2010. Here we provide the following original documents packaged as one in the following sequence:

1. the initial public summary document of the FBE released first on 2 Nov. 2009 with 3 candidate bubbles and updated with 1 more candidate bubble on 23 Dec. 2009, which includes the digital fingerprints of the four original separate documents (one for each asset) reproduced at the end;

2. the final analysis of the four bubbles released on 3 May 2010;

3. the four original documents fbe_001.pdf, fbe_002.pdf, fbe_003.pdf prepared on 2 Nov. 2009, and fbe_004.pdf prepared on 22 Dec. 2009.

For the purpose of verifying the checksums of the four individual original documents, they and the rest of the contents of this summary document can be found online at:
`http://www.er.ethz.ch/fco/index`.

*Electronic address: dsornette@ethz.ch

# The Financial Bubble Experiment: advanced diagnostics and forecasts of bubble terminations

D. Sornette, R. Woodard, M. Fedorovsky, S. Reimann, H. Woodard, W.-X. Zhou
(The Financial Crisis Observatory)*
*Department of Management, Technology and Economics,*
*ETH Zurich, Kreuzplatz 5, CH-8032 Zurich, Switzerland*

On 2 November 2009, the Financial Bubble Experiment was launched within the Financial Crisis Observatory (FCO) at ETH Zurich (`http://www.er.ethz.ch/fco/`). In that initial report, we diagnosed and announced three bubbles on three different assets. In this latest release of 23 December 2009 in this ongoing experiment, we add a diagnostic of a new bubble developing on a fourth asset.

## I. INTRODUCTION

The motivation of the Financial Bubble Experiment finds its roots in the failure of standard approaches. Indeed, neither the academic nor professional literature provides a clear consensus for an operational definition of financial bubbles or techniques for their diagnosis in real time. Instead, the literature reflects a permeating culture that simply assumes that any forecast of a bubble's demise is inherently impossible.

The Financial Bubble Experiment (FBE) aims at testing the following two hypotheses:

- **Hypothesis H1:** financial (and other) bubbles can be diagnosed in real-time before they end.
- **Hypothesis H2:** The termination of financial (and other) bubbles can be bracketed using probabilistic forecasts, with a reliability better than chance (which remains to be quantified).

Because back-testing is subjected to a host of possible biases, we propose the FBE as a real-time advanced forecast methodology that is constructed to be free, as much as possible, of all possible biases plaguing previous tests of bubbles. In particular, active researchers are constantly tweaking their procedures, so that predicted 'events' become moving targets. Only advance forecasts can be free of data-snooping and other statistical biases of ex-post tests. The FBE aims at rigorously testing bubble predictability using methods developed in our group and by other scholars over the last decade. The main concepts and techniques used for the FBE have been documented in numerous papers (Jiang et al., 2009; Johansen et al., 1999; Johansen and Sornette, 2006; Sornette and Johansen, 2001; Sornette and Zhou, 2006) and the book (Sornette, 2003). The FCO research team is currently developing and testing novel estimations methods that will be progressively implemented in future releases.

In the FBE, we propose a new method of delivering our forecasts where the results are revealed only after the predicted event has passed but where the original date when we produced these same results can be publicly, digitally authenticated.

Since our science and techniques involve forecasting, the best test of a forecast is to publicize it and wait to see how accurate it is, whether the wait involves days, weeks or months (we rarely make forecasts for longer time scales). While we fully intend to do this, we want to delay the unveiling of our results until after the forecasted event has passed to avoid potential issues of liability, ethics and speculation. Also, we think that a full set of results showing multiple forecasts all at once is more revealing of the quality of our current methods than would be a trickle of one such forecast every month or so. We also want to address the obvious criticism of cherry picking successful forecasts, as explained below. In order to be convincing, our experiment has to report all cases, be they successes or failures.

## II. DESCRIPTION OF THE METHODOLOGY OF THE FINANCIAL BUBBLE EXPERIMENT

Our proposed method for this experiment is the following:

---

*Electronic address: `dsornette@ethz.ch`

- We choose a series of dates with a fixed periodicity on which we will reveal our forecasts and make these dates public by immediately posting them on our University web site and on the first version of our main publication, which we describe below. Specifically, our first publication of the forecasts will be issued on 1 May 2010 and then in successive deliveries every 6 months. However, we keep open the option of changing the periodicity of the future deliveries as the experiment unfolds and we learn from it and from feedback of the scientific community.
- We then continue our current research involving analysis of thousands of global financial time series.
- When we have a confident forecast, we summarize it in a simple document, such as a .pdf file that is easily viewed by almost all modern desktop and laptop computers.
- We do not make this document public. Instead, we make its digital fingerprint public. We generate three digital fingerprints for each document, with the publicly available (1) MD5 hash algorithm [1] and (2) 256 and 512 bit versions of the SHA-2 hash algorithm [2] [3]. This creates three strings of letters and numbers that are unique to this file. Any change at all in the contents of this file will result in different MD5 and SHA-2 signatures.
- We create the first version of our main document, containing the first two sections of this document, a brief description of our theory and methods, the MD5 and SHA-2 hashes of our first forecast and the date (1 May 2010) on which we will make the first original .pdf document public.
- We upload this main 'meta' document to `http://arxiv.org`. This makes public our experiment and the MD5 and SHA-2 hashes of our first forecast. In addition, it generates an independent timestamp documenting the date on which we made (or at least uploaded) our forecast. `arxiv.org` automatically places the date of when the document was first placed on its server as 'v1' (version 1). It is important for the integrity of the experiment that this date is documented by a trusted third party.
- We continue our research until we find our next confident forecast. We again put the forecast results in a .pdf document and generate the MD5 and SHA-2 hashes. We now update our master document with the date and digital fingerprint of this new forecast and upload this latest version of the master document to `arxiv.org`. The server will call this 'v2' (version 2) of the same document while keeping 'v1' publicly available as a way to ensure integrity of the experiment (i.e., to ensure that we do not modify the MD5 and SHA-2 hashes in the original document). Again, 'v2' has a timestamp created by `arxiv.org`.
- Notice that each new version contains the previous MD5 and SHA-2 signatures, so that in the end there will be a list of dates of publication and associated MD5 and SHA-2 signatures.
- We continue this protocol until the future date (1 May 2010) at which time we upload our final version of the master document. For this final version, we include the URL of a web site where the .pdf documents of all of our past forecasts can be downloaded and independently checked for consistent MD5 and SHA-2 hashes. For convenience, we will include a summary of all of our forecasts in this final document.

Note that the above method implies two aspects of the same important check to the integrity of our experiment:

1. We will reveal all forecasts, be they successful or not.
2. We will not simply 'cherry-pick' the results that we would want the community to see (with a few token, possibly, bad results). We do not have another simultaneous outlet where we are running a similar experiment, since `arxiv.org` is a very visible international platform.

Once the .pdf documents with the full description of the forecasts are made public, the question arises as how to evaluate the quality of the diagnostics and how these results help falsify the two hypotheses? In a nutshell, the problem boils down to qualifying (and quantifying) what is meant by (i) a successful diagnostic of the existence of a bubble and (ii) a successful forecast of the termination of the bubble. In the end, one would like to develop statistical tests to falsify the two hypotheses stated above, using the track record that the present financial bubble experiment has the aim to construct. We leave these issues for future work and discussions, while we realize that one would have liked in principle to state a precise definition of successes. For instance, Chapter 9 of (Sornette, 2003) suggests a number of options, including the "statistical roulette", Bayesian inference and error diagrams. Our main goal with this FBE is to timestamp our forecasts as we simultaneously continue our search for adequate measures to qualify the quality of our forecasts.

## III. BACKGROUND AND THEORY

Our theories of financial bubbles and crashes have been well-researched and documented over the past 15 years in many papers and books. We refer the reader to the Bibliography. In particular, broad overviews can be found in (Johansen et al., 1999; Johansen and Sornette, 2006; Sornette, 2003; Sornette and Johansen, 2001; Sornette and Zhou, 2006). In short, our theories are based on positive feedback on the growth rate of an asset's price by price, return and other financial and economic variables, which lead to faster-than-exponential (power law) growth. The positive feedback is partially due to imitation and herding among humans, who are actively trading the asset. This signature is quantitatively identified in a time series by a faster-than-exponential power law component, the existence of increasing low-frequency volatility, these two ingredients occurring either in isolation or simultaneously with varying relative amplitudes. A convenient representation has been found to be the existence of a power law growth decorated by oscillations in the logarithm of time. The simplest mathematical embodiment is obtained as the first order expansion of the log-periodic power law (LPPL) model and is shown in Eq. (1):

$$\ln P = A + B|t - t_c|^{\alpha} + C|t - t_c|^{\alpha} \cos[\omega \ln |t - t_c| + \phi] \quad (1)$$

where $P$ is the price of the asset and $t$ is time. There are 7 parameters in this nonlinear equation. Our past work has led to the hypothesis that the LPPL signals can be useful precursors to an ending (change of regime) of the bubble, either in a crash or a less-dramatic leveling off of the growth.

## IV. METHODS

As are our theories, our methods are documented elsewhere so we only briefly mention the general technique so that the forecasts that we make public can be better understood. In short, we scan thousands of financial time series each week and identify regions in the series that are well-fit by Eq. (1). We divide each time series into sub-series defined by start and end times, $t_1$ and $t_2$ and then fit each sub-series $(t_1, t_2)$. We choose $\max(t_2)$ as the date of the most recent available observation and $\min(t_2)$ as 31 days before. Many sub-series are created according to the following parameters: $dt_1 = dt_2 = 10$ days, $\min(t_2 - t_1) = 110$ days and $\max(t_2 - t_1) = 1500$ days. In practice, this maximum series length is far longer than is relevant, but we use it to ensure capturing the beginning of a bubble regime in our parameter mesh.

After filtering all fits with an appropriate range of parameters, we select those assets that have the strongest LPPL signatures. We calculate the residues between the model and the observations and use the residues to create 10 synthetic datasets (bootstraps) that have similar statistics as the original time series. We fit Eq. (1) to the synthetic data and then extrapolate this entire ensemble of LPPL models to six months beyond our last observation. One of the parameters in the LPPL equation is the "crash" time $t_c$, which represents the most probable time of the end of the bubble and change of regime. We identify the 20%/80% and 5%/95% quantiles of $t_c$ of the fits of the extrapolated ensemble that have $t_c$ within this six month range. These two sets of quantiles, the date of the last observation and the number of fits in the ensemble are published in our forecasts.

## V. BUBBLE FORECASTS

Table I lists the checksums of our forecast documents. Each line (and, therefore, each checksum) represents a different, distinct forecast on a separate time series.

The first release of this document was on 2 November 2009. As there was concern about the vulnerability of the MD5 algorithm, we updated this document on 6 November 2009 with the corresponding SHA-2 256 and 512 checksums for each document. We leave the MD5 checksum as is for each document. The following checksums were made with the `md5sum`, `sha256sum` and `sha512sum` programs on a GNU/Linux machine. The following output shows the version of the programs used:

```
$ md5sum --version
md5sum (GNU coreutils) 6.10
Copyright (C) 2008 Free Software Foundation, Inc.
License GPLv3+: GNU GPL version 3 or later <http://gnu.org/licenses/gpl.html>
This is free software: you are free to change and redistribute it.
There is NO WARRANTY, to the extent permitted by law.

Written by Ulrich Drepper, Scott Miller, and David Madore.

$ sha256sum --version
sha256sum (GNU coreutils) 6.10
Copyright (C) 2008 Free Software Foundation, Inc.
License GPLv3+: GNU GPL version 3 or later <http://gnu.org/licenses/gpl.html>
This is free software: you are free to change and redistribute it.
There is NO WARRANTY, to the extent permitted by law.

Written by Ulrich Drepper, Scott Miller, and David Madore.

$ # 'sha512sum --version' gives same information

$ shasum --version
5.45
```

## VI. LINK TO FORECAST DOCUMENTS

To be written on 1 May 2010.

## VII. CONCLUSIONS

To be written after 1 May 2010.

| Publication date<br>Document name | MD5SUM<br>SHA256SUM<br>SHA512SUM |
|---|---|
| 2009-11-02<br>fbe_001.pdf | 6d9479eb2849115a12c219cfa902990e<br>d7ad5c9531166917ba97f871fb61bd1f6290b4b4ce54e3ba0c26b42e2661dc06<br>808bbfaddbca3db8d0f55d74cabedf5201ecd70340f86e27dfac589ce682144f52f6fc4b3ff1ac75231038d86dae58bd320e7fb17ef321b4bc61a19e88071039 |
| 2009-11-02<br>fbe_002.pdf | 5d375b742a9955d4aeea1bd5c7220b2b<br>5a9c395b9ab1d2014729ac5ff3bb22a352e14096fa43c59836ea0d4ae0e3b453<br>e7ef9150b4738253f4021b0600eff1cd455b2671e421b788b9268b518439b56699994b3f8b395742bdc7622b5536034e74ade86e0a46bff71ed5ff9a293f809f |
| 2009-11-02<br>fbe_003.pdf | fd85000d0ce3231892ef1257d2f7ab1e<br>d3f3d504d85d50eb3dc0fe2c3042746db2f010509f4d1717370d14012972e86f<br>91a8fa82b7f08deea2df2a1f7cef266f5aa155bb0c047f65b14315f7229d92976cc7b30453453fb8ecd0350783907c83652192d32ba90fb1cce128385832e63a |
| 2009-12-23<br>fbe_004.pdf | 8e019304004ebf06df17384ff664ff57<br>27c650d85a802eafecd8389391c440458816ff13b5c573bab710e3b7739f2e38<br>388fa7941c691fe7c8887886a932dd6a6aa28a967b5b05bf3cf96cdb836b499f354a78bca67d86aa246985b80e75670c3bd6300f6f4f92ca3bd0b59ac675e1eb |

TABLE I: Checksums of Financial Bubble Experiment forecast documents. The documents fbe_001.pdf to fbe_004.pdf correspond to diagnostics of 4 different bubbles in 4 distinct assets.

5

# The Financial Bubble Experiment
# First Results (2 November 2009 - 1 May 2010)

D. Sornette, R. Woodard, M. Fedorovsky, S. Reimann, H. Woodard, W.-X. Zhou (The Financial Crisis Observatory)*
*Department of Management, Technology and Economics,*
*ETH Zurich, Kreuzplatz 5, CH-8032 Zurich, Switzerland*

On 2 November 2009, the Financial Bubble Experiment was launched within the Financial Crisis Observatory (FCO) at ETH Zurich (`http://www.er.ethz.ch/fco/`). In that initial report, we diagnosed and announced three bubbles on three different assets (IBOVESPA Brazil Index, a Merrill Lynch corporate index, gold spot price). In the subsequent release of 23 December 2009 in the ongoing experiment, we added a diagnostic of a new bubble developing on a fourth asset (cotton futures). This present report presents the four initial forecasts and analyses how they fared. We find that IBOVESPA and gold showed clear signs of changing from a bubble regime to a new one within our forecast quantile windows; that the Merrill Lynch bond index changed from a strong bubble regime to one of more moderate growth just before our publication date; and that cotton was and still is in a bubble without showing a clear change of regime.

## I. INTRODUCTION

The Financial Bubble Experiment (FBE) aims at testing the following two hypotheses:

- **Hypothesis H1:** Financial (and other) bubbles can be diagnosed in real-time before they end.
- **Hypothesis H2:** The termination of financial (and other) bubbles can be bracketed using probabilistic forecasts, with a reliability better than chance (which remains to be quantified).

In [1], we described the methodology of the FBE, the background and underlying theory as well as the practical procedure. Three forecasts summarized in three `.pdf` documents were documented in the first release of the document on 2 Nov. 2009. A second version was released on 6 Nov. 2009 adding the SHA-2 256 and 512 checksums to the initial MD5 check sums for each of the three documents to address the concern about the vulnerability of the MD5 algorithm. A fourth forecast was added in the third version of the document on 23 Dec. 2009. Refer to Table V.

Our ambition in this first report is to diagnose bubbles in the time frame of our experiment: we aim at diagnosing and forecasting bubbles in various major asset classes which have the potential to end within a window of about 6 months or less (which is the duration between each successive release of the FBE). Because in a population of bubbles of various amplitudes, the moderate sized ones will be much more numerous that the great ones, the moderate sized bubbles are likely to be represented more often in our sample population. We provide evidence of clear corrections and *changes of regimes* instead of reports of crashes of very major amplitudes.

Ideally, we would like to aim at diagnosing in advance and predicting the turning point of major bubbles, which have the potential for devastating crashes. Examples of such bubbles include those that ended in the Oct. 1987, March-April 2000 and 2007-2008 crashes. Such events are relatively rare, typically recurring once or twice per decade at most. It would be unrealistic to specifically aim at such rare events in our experiment, because the time scale needed to reach any reliable statistical conclusion would be too long in relation to, say, an academic career. As we continue this experiment, we will almost certainly encounter such monsters.

## II. LINK TO FORECAST DOCUMENTS

The original `.pdf` documents can be found via links at the FCO web page at `http://www.er.ethz.ch/fco/`.

*Electronic address: `dsornette@ethz.ch`

## III. EXECUTIVE SUMMARY OF THE RESULTS

The four assets (and their datasources) that we identified are:

- IBOVESPA Index, Brazil (Yahoo: ^BVSP)
- Merrill Lynch EMU (European Monetary Union) Corporates Non-Financial Index (Bloomberg: EN00)
- Gold spot price in USD/oz (Bloomberg: GOLDS Comdty)
- Cotton future in USD/pound (Bloomberg: CT1 COMB Comdty)

We summarize the quantitative results in the following sections. A complete visual representation is available in the figures in Section X.

### H1: Identification of a bubble

1. We support H1 by confirming that bubbles existed in two of the four assets at the time of our forecasts ($t_2 = 2$ Nov. 2009 release for IBOVESPA and gold). Our more recently developed 'bubble index' (Section IX) confirms that IBOVESPA is out of its previous bubble regime, thus confirming the existence of the announced changed of regime.
2. We also confirm that a bubble in the Merrill Lynch Index did exist, but it ended just before our forecast date. This is clear in retrospect with the insight of the 6 additional months of data. Our more recently developed 'bubble index' (Section IX) confirms that the Merrill Lynch index was just exiting a bubble at our forecast time. Thus, it is now firmly out of a bubble regime.
3. For cotton (23 Dec. 2009 release), the metrics given in section VIII give ambiguous signals concerning a change of regime. Our more recently developed 'bubble index' (Section IX) diagnoses that the bubble is still continuing. We cannot therefore conclude yet on the validity of our diagnostic as we have to wait on its subsequent development.

Sections V-VIII present standard measures of regime shifts performed with the benefit of the full price time series through the end of April.

### H2: Forecast of change of regime

1. IBOVESPA and gold support H2 by clearly exhibiting changes of regime beginning within our forecasted quantile windows.
2. It is clear now that the change of regime in the Merrill Lynch Index started just before our forecast date. We note that the early limit for both of our quantile windows for this asset included dates before we made the forecast. That is, our analysis indicated that it was possible that the change had already begun, though that only becomes clear with the trends seen in the full time series.
3. For cotton, the quantile range predicted correctly the 12% drawdown occurring right its middle. This is however a partial success (or failure) for H2, since our present indicators diagnose that the bubble has not ended, suggesting that we previously identified a "baby" bubble (with a significant accident) that is still growing into a larger bubble.

**Brazil IBOVESPA (`fbe_001.pdf`):** Change of regime began within our forecast window: large drawdown of 11% in 30 days occurred approximately two weeks after the end of our largest forecast window. Peak and subsequent decline of fraction of positive return days and sharp drop in local growth rate of price occurred within forecast window.

| | **Start** | **End** |
|---|---|---|
| 05/95 | 2009-10-19 | 2009-12-17 |
| 20/80 | 2009-10-27 | 2009-11-29 |

TABLE I: Published forecast quantile ranges for IBOVESPA.

**Merrill Lynch bond index (fbe_002.pdf):** Change of regime began 1-2 months before our publication date: price trend changed from sharp rise (20% per year) over almost one year (Nov. 2008 - Sept. 2009) to a more moderate rise (9.1% per year). Fraction of positive return days and local growth rate confirm the change began just before our forecast date.

| | **Start** | **End** |
|---|---|---|
| 05/95 | 2009-10-11 | 2010-02-09 |
| 20/80 | 2009-10-27 | 2010-01-16 |

TABLE II: Published forecast quantile ranges for Merrill Lynch bond index.

**Gold spot price (fbe_003.pdf):** Change of regime began well within our 20-80% forecast window: large drawdown of 13% in 68 days (which includes an initial drop of 11% in 20 days). Positive return days and local growth rate confirm the timing of this transition.

| | **Start** | **End** |
|---|---|---|
| 05/95 | 2009-10-13 | 2010-09-07 |
| 20/80 | 2009-11-05 | 2010-02-25 |

TABLE III: Published forecast quantile ranges for gold spot price.

**Cotton futures (fbe_004.pdf):** Drawdown of 12% in 32 days began within our forecast window, though, it is comparable to other drawdowns in the previous year. This drawdown was followed by a steep run-up that has now plateaued. Fraction of positive return days suggest a change of regime but local growth rate analysis is inconclusive.

| | **Start** | **End** |
|---|---|---|
| 05/95 | 2009-12-05 | 2010-04-09 |
| 20/80 | 2009-12-31 | 2010-03-16 |

TABLE IV: Published forecast quantile ranges for cotton futures.

## IV. ANALYSIS OF THE FOUR ASSETS

In the following subsections, we include figures for each asset class. The figures share some common features, explained here. Details for each individual asset and measure are discussed below.

**Axes and observations:** Price (or value of the asset) is shown on the left vertical axis and calendar days are indicated on the bottom horizontal axis. The blue circles represent closing price observations on trading days. Note that the price axis shows observations in their natural units on a logarithmic scale. Data for IBOVESPA was obtained from Yahoo financial for symbol '^BVSP' (using adjusted close) and data for the other three assets (Merrill Lynch bond index, gold spot price and cotton future) were obtained from Bloomberg.

**Large shaded region and vertical black line:** The large grey shaded regions that begin near the left price axis and end at the solid black vertical line near the right vertical axis represent the domain of the observations used in our analysis. The vertical line itself sits at $t_2$: the last observation used in the analysis.

**Small shaded regions:** Two colored shaded regions begin in the vicinity of $t_2$. They represent our forecast "danger" zones, where changes of regimes are most likely to occur. The inner, narrow, green one with horizontal hatching represents the 20-80% quantile interval and the outer, wider orange (without hatching) represents the 5-95% quantile interval. That is, these two numbers imply a 60% (respectively, 90%) probability for the end of the bubble to be located within the green (respectively, orange) zone.

**Zoomed figures:** Some figures below show a zoomed-in version of a main figure, where the vertical line at $t_2$ is centered.

**Drawdown analysis:** Drawdown analysis figures show a solid red line connecting the path of the largest drawdown observed between $t_2$ and the most recent observations used in this document (27 April 2010). The percentage drop and duration of this drawdown is indicated in text in the lower right corner of each figure. A drawdown is simply defined as the largest peak-to-trough drop in price in a given region.

**Fraction of up days in a running window:** We calculate one day close-to-close returns for each asset and mark them as positive (up) or non-positive (zero or down). The ratio of up days relative to the sum of up and down days in a running window of 30, 60 or 90 days is plotted on top of the price observations. The right vertical axis shows this fraction on a logarithmic scale. Note that we do not include returns with a value of zero in the calculation of this ratio. Also, the running window *ends* at the value plotted on the time axis. That is, only present and past data is used in the running window, not future data.

**Derivative of observations:** Another measure of the change of regime is provided by an estimation of the local growth rate. We use the Savitzky-Golay smoothing algorithm to calculate the first derivative of the observations, using a third order polynomial fit centered within windows of 120 and 180 days. The scale of the estimated derivative is shown on the right vertical axis using linear scaling.

## V. ANALYSIS OF THE BRAZIL INDEX (IBOVESPA)

Figure 1 summarizes the evolution of the Brazil index (IBOVESPA) over the period of interest. Figure 2 provides a zoomed view of the same information. Visual inspection suggests that indeed there has been a change of regime, coincident to the danger zones delineated in the two shaded quantile regions. Here the change of regime corresponds to a transition from a fast ascending price to a noisy plateau without growth.

The existence of the change of regime occurring roughly in the danger zones is made perhaps more quantitative by Figure 3 showing the fraction of days with positive returns (close-minus-close) in a moving window of 90 days. One can observe that this fraction was continuously increasing over the interval represented in grey over which we identified a bubble, culminating at the value of 65% exactly in the middle of the 5-95% danger zone. This peak was followed by a steep decline back to a no-trend value of close to 50%.

Figure 4 shows that the two estimations of the local growth rate plunged to close to zero just in the danger zones, confirming the visual impression of a change of regime.

## VI. ANALYSIS OF THE MERRILL CORP. NON-FINANCIAL INDEX

Figure 5 summarizes the evolution of the Merrill Lynch EMU (European Monetary Union) Corporates Non-Financial Index over the periods of interest. There has been a change of regime, in the sense that the growth rate has rather brutally changed in August-September 2009, as indicated by the two trends shown as the two straight lines. The leftmost, steeper trend line has a slope equivalent to an annualized return of 20% while that of the rightmost, flatter trend line is 9.1%. But, of course, such a change in trend is only clear in retrospect when sufficient data has been accumulated to confirm it. The change was not clear at the time when our forecast was performed. Our method has identified this change of trend occurring after a strong growth regime as the signal of a more dramatic bubble end.

Figure 6 shows the fraction of days with positive returns (close-minus-close) in a moving window of 90 days. One can observe that this fraction has been continuously high, oscillating around 70% for most of 2009, until it started to descend below 60% over the time interval indicated by the danger zones.

Figure 7 shows that the estimations of the local growth rate of the Merrill Index in two moving windows (60 and 90 days) plunged to close to zero slightly before the danger zones, confirming the visual impression of the change of growth regime.

## VII. ANALYSIS OF GOLD

Figures 8 and 9 summarize the evolution of the gold spot price over the period of interest. Visual inspection suggests that indeed there has been a change of regime, coincident with the danger zones delineated in the two quantile windows. Here the change of regime corresponds to the transition from a fast ascending price to a significant drawdown followed by a noisy plateau without growth. Note that the drawdown of 13% in 68 days includes an initial steep drawdown of 11% in 20 days.

Figure 10 shows the fraction of days with positive returns (close-minus-close) in a moving window of 90 days. One can observe that this fraction was increasing steadily since the end of 2008, peaking at a value of 64% exactly in the middle of the 20-80% danger zone. This peak was followed by a steep decline back to a no-trend value of close to 50%.

The Savitzky-Golay smoothed local estimations of the growth rate of the Gold spot price in 60 and 90 day moving windows show drops into negative values just in the danger zones, confirming the visual impression of a change of regime.

## VIII. ANALYSIS OF COTTON FUTURE PRICE IN USD

Figure 12 summarizes the evolution of the cotton future price in USD over the periods of interest. It is not clear that there has been a change of regime. One can see a drawdown outlined in red with a cumulative loss of 12% in 32 days, but it is followed by a strong rebound, almost recovering the previous trend. Visual inspection may be misleading and we resort to the previous usual metrics.

Figure 13 shows the fraction of days with positive returns (close-minus-close) in a moving window of 90 days. One can observe that this fraction has been increasing steadily and then plateaued over the interval indicated by the grey zone over which we diagnosed a bubble, culminating at a value of almost 65% in the first tier part of the 20-80% and 5-95% danger zones. This peak was followed by a clear decline developing over the danger zones.

Figure 14 shows an estimate of the local growth rate using the same parameters as for the three other cases. This metric shows no change of regime.

## IX. WHAT WOULD WE DO DIFFERENTLY NOW?

Since the first incarnation of the FBE implemented in the first version of this document [1], monitoring the development of the forecast in real time has been quite a learning experience. Several improvements have been developed which will be incorporated into the next set of forecasts. Here, we provide just a flavor by showing a new metric and the impact it would have had on these four forecasts.

The metric is a "bubble index", which quantifies the probability for a bubble to be present and is shown for each asset in Figures 15- 18. This metric mainly aims at hypothesis H1, the diagnosis that a bubble is present. As can be seen in Figure 16, we would not have chosen the forecast for Merrill Corp. Non-Financial Index because the bubble index was telling us loudly and clearly that the bubble had formed well in advance of our publication date and had already ended when we made our forecast. This new bubble index supports our previous analysis above of diagnosing bubbles in IBOVESPA and gold, with high values of the index leading up to and into the quantile regions. Also, this metric and others show a clear signal for cotton, supporting our claim that this asset is still in a bubble regime.

| Publication date<br>Document name<br>Asset | MD5SUM<br>SHA256SUM<br>SHA512SUM |
|---|---|
| 2009-11-02<br>fbe_001.pdf<br>IBOVESPA (Brazil) | 6d9479eb2849115a12c219cfa902990e<br>d7ad5c9531166917ba97f871fb61bd1f6290b4b4ce54e3ba0c26b42e2661dc06<br>808bbfaddbca3db8d0f55d74cabedf5201ecd70340f86e27dfac589ce682144f52f6fc4b3ff1ac75231038d86dae58bd320e7fb17ef321b4bc61a19e88071039 |
| 2009-11-02<br>fbe_002.pdf<br>ML Corp. Non-Fin. Index | 5d375b742a9955d4aeea1bd5c7220b2b<br>5a9c395b9ab1d2014729ac5ff3bb22a352e14096fa43c59836ea0d4ae0e3b453<br>e7ef9150b4738253f4021b0600eff1cd455b2671e421b788b9268b518439b56699994b3f8b395742bdc7622b5536034e74ade86e0a46bff71ed5ff9a293f809f |
| 2009-11-02<br>fbe_003.pdf<br>Gold spot price (USD) | fd85000d0ce3231892ef1257d2f7ab1e<br>d3f3d504d85d50eb3dc0fe2c3042746db2f010509f4d1717370d14012972e86f<br>91a8fa82b7f08deea2df2a1f7cef266f5aa155bb0c047f65b14315f7229d92976cc7b30453453fb8ecd0350783907c83652192d32ba90fb1cce128385832e63a |
| 2009-12-23<br>fbe_004.pdf<br>Cotton futures (USD) | 8e019304004ebf06df17384ff664ff57<br>27c650d85a802eafecd8389391c440458816ff13b5c573bab710e3b7739f2e38<br>388fa7941c691fe7c8887886a932dd6a6aa28a967b5b05bf3cf96cdb836b499f354a78bca67d86aa246985b80e75670c3bd6300f6f4f92ca3bd0b59ac675e1eb |

TABLE V: Checksums of Financial Bubble Experiment forecast documents. The documents fbe_001.pdf to fbe_003.pdf contain the diagnostics of 3 different bubbles in 3 distinct assets, and were finalized on 2 Nov. 2009. The document fbe_004.pdf contains the diagnostic of an additional bubble of another asset, which was finalized on 23 Dec. 2009. This table was released on 2 Nov. 2009 and updated on 23 Dec. 2009 in the document [1]. Here, we add the name of the asset corresponding to each document.

## X. FIGURES

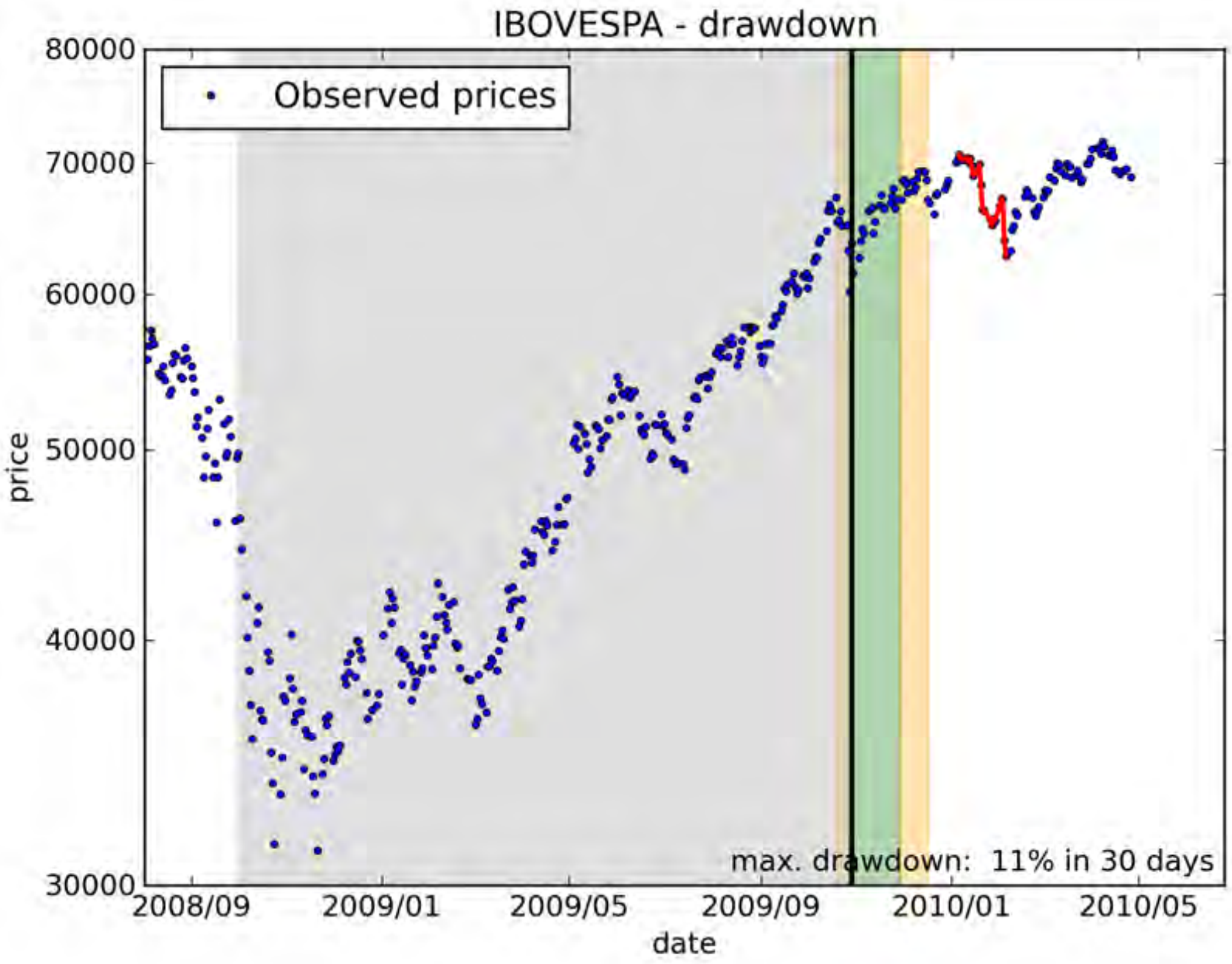

FIG. 1: Time series of the Brazil Index (IBOVESPA) shown as filled blue circles, the 20-80% (respectively 5-95%) quantile intervals for the predicted end of the bubble and the subsequent evolution of the index. The black vertical line shows the time of the last observation used in the analysis. The red solid trace shows the largest drawdown that occurred after the forecast, 11% in 30 days.

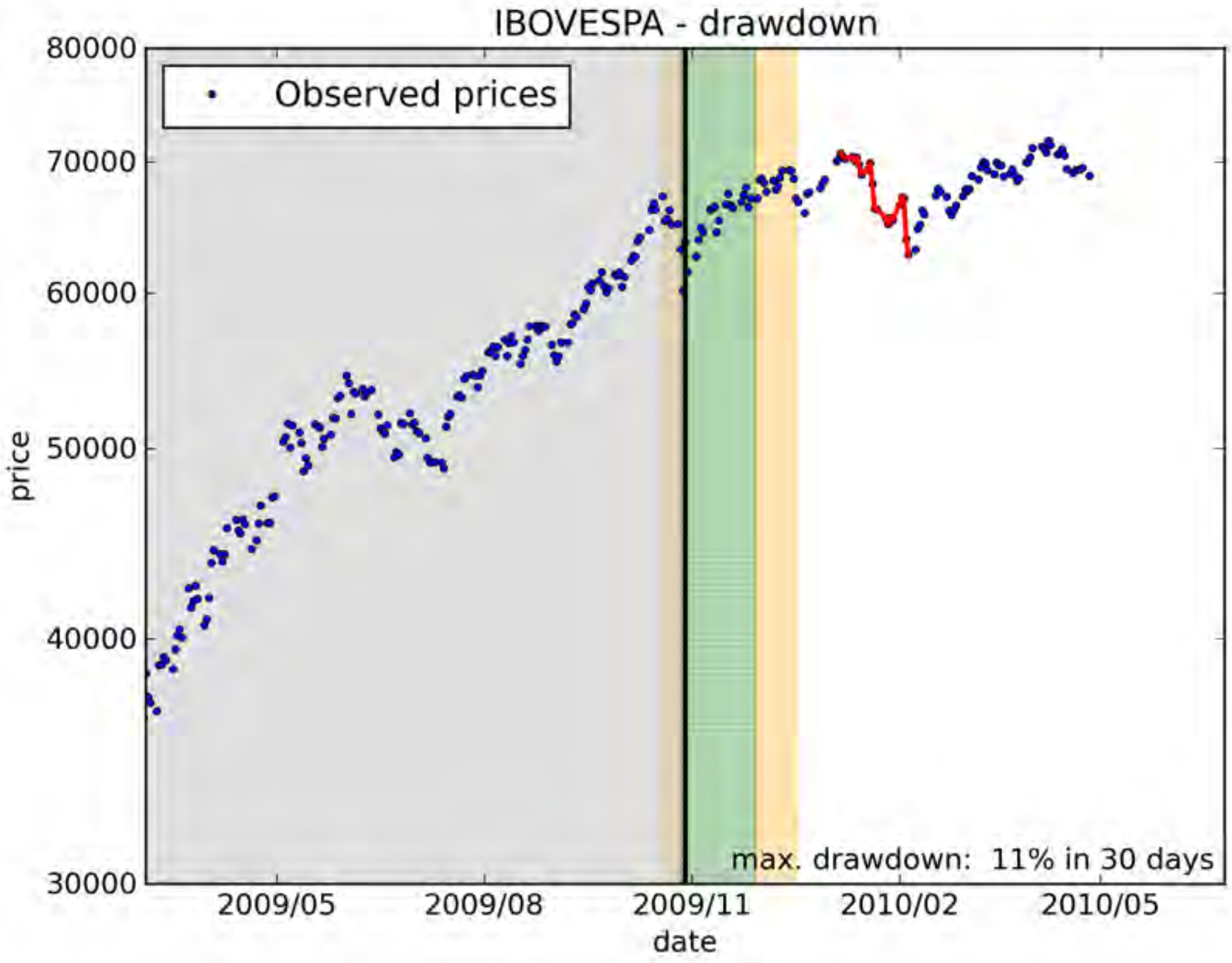

FIG. 2: Zoom of Figure 1.

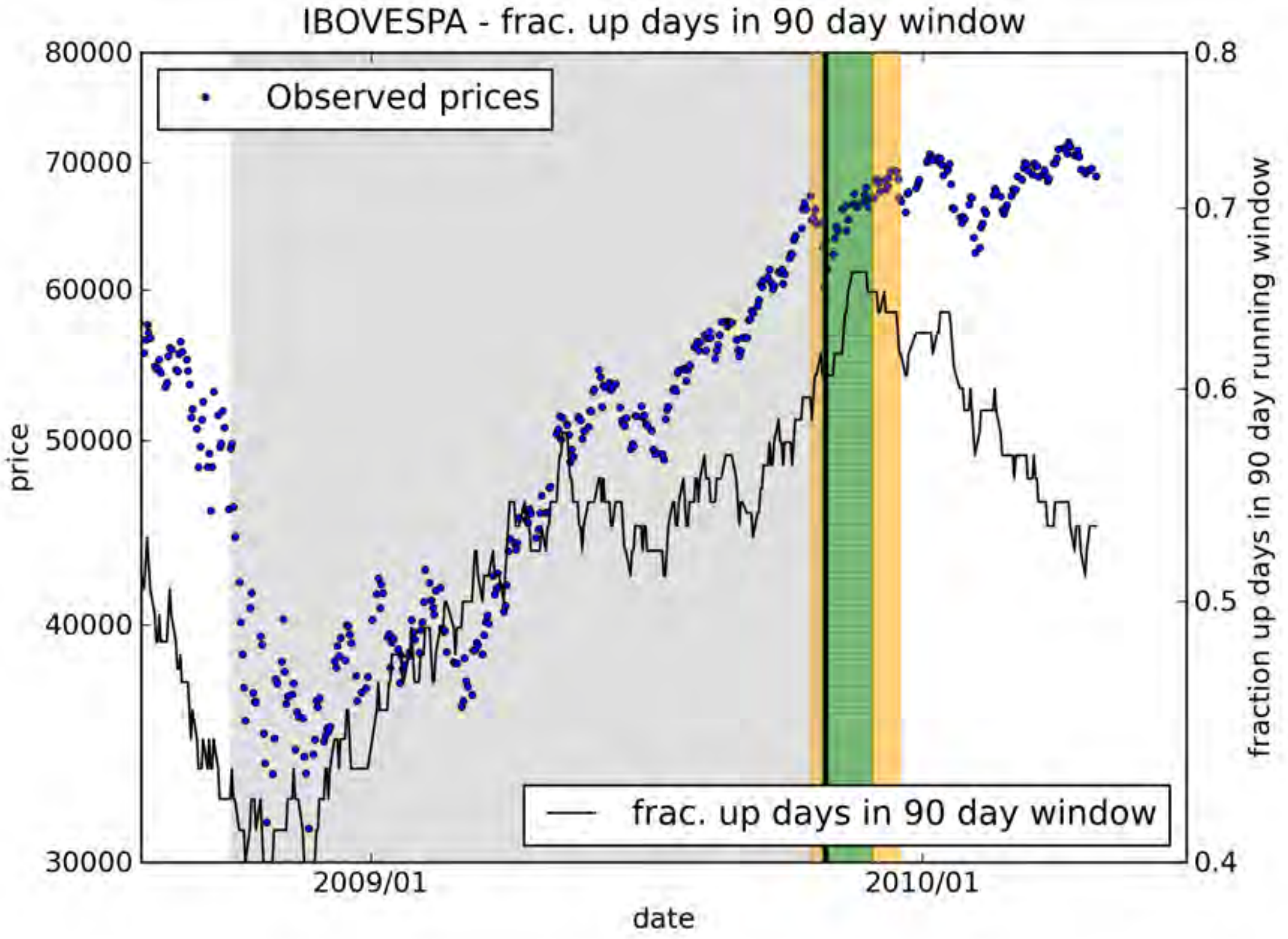

FIG. 3: On the data shown in Figure 1, we plot the fraction of days (right vertical scale) with positive returns as a function of the right-end time of a moving window of width equal to 90 days.

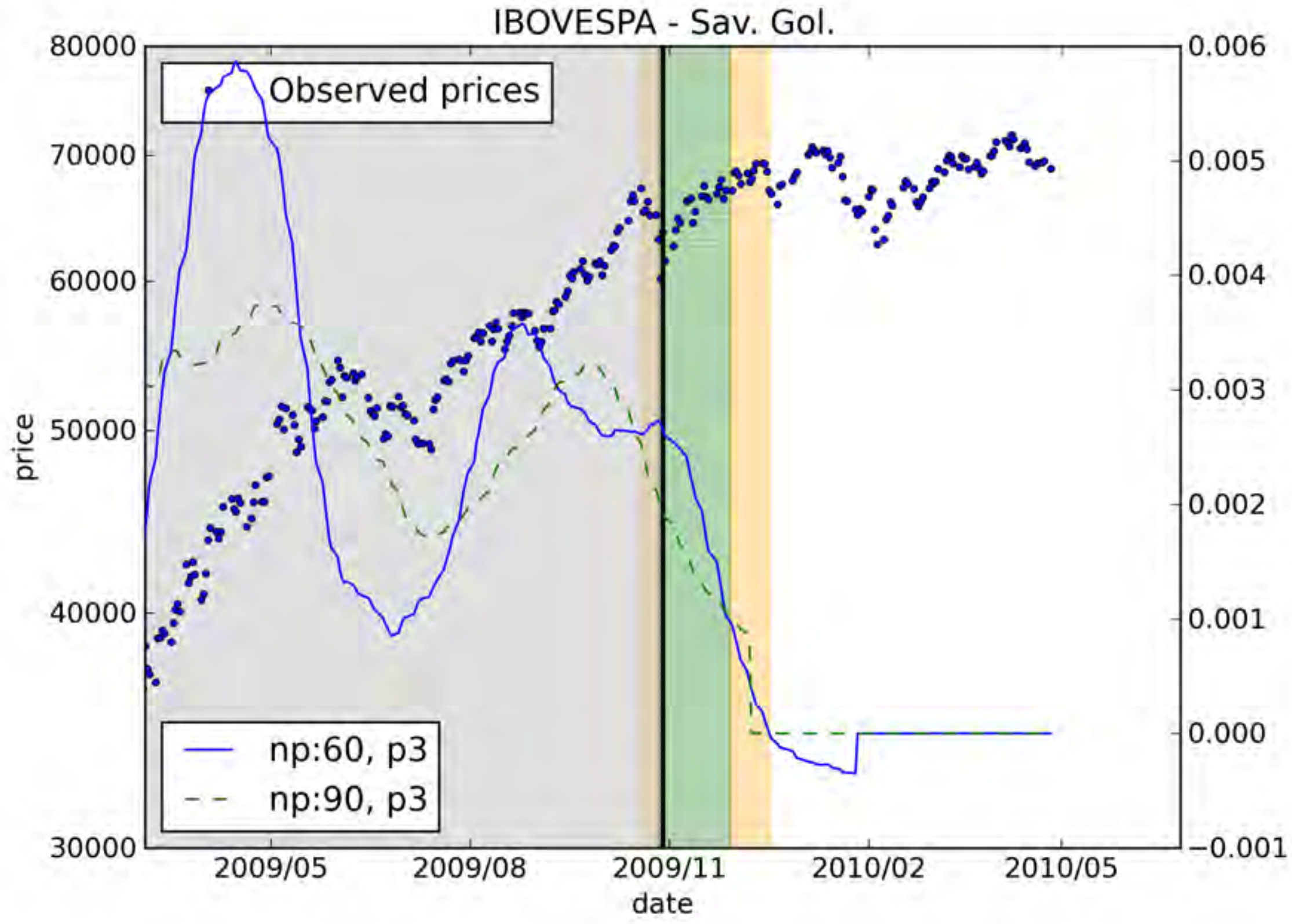

FIG. 4: On the data shown in Figure 1, the dashed green line (respectively continuous blue line) gives the smoothed growth rate of the Brazil index in the middle of a running window of 180 (respectively 120 days), as estimated with the Savitzky-Golay smoothing algorithm with a polynomial of order 3.

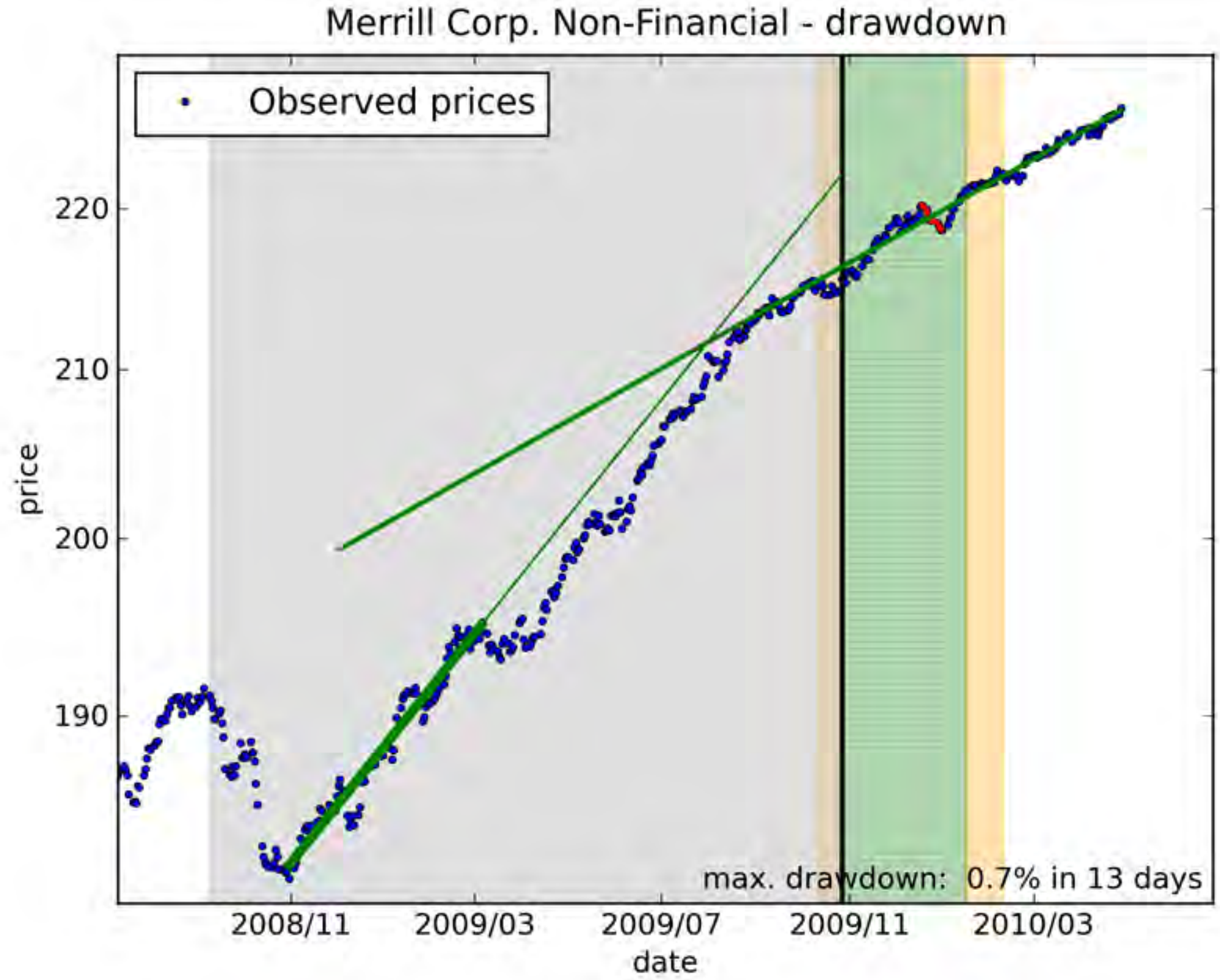

FIG. 5: Time series of the Merrill Lynch EMU (European Monetary Union) Corporates Non-Financial Index (Total Return Index, EN00) shown as filled blue circles, the 20-80% (respectively 5-95%) quantile intervals for the predicted end of the bubble and the subsequent evolution of the index. The black vertical line shows the time of the last observation used in the analysis. The red solid trace shows the largest drawdown that occurred after the forecast, 0.7% in 13 days.

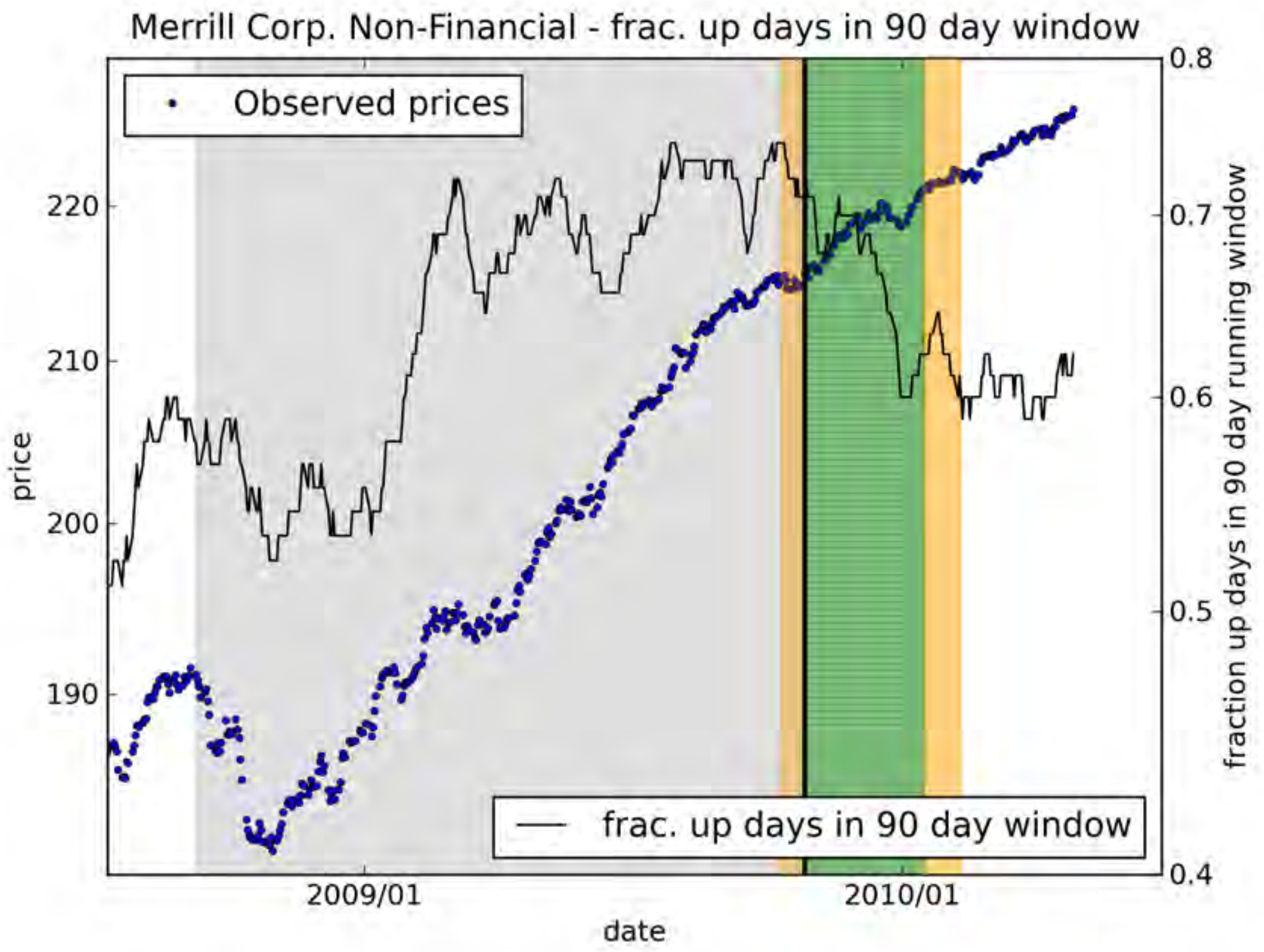

FIG. 6: On the data shown in Figure 5, we plot the fraction of days (right vertical scale) with positive returns as a function of the right-end time of a moving window of width equal to 90 days.

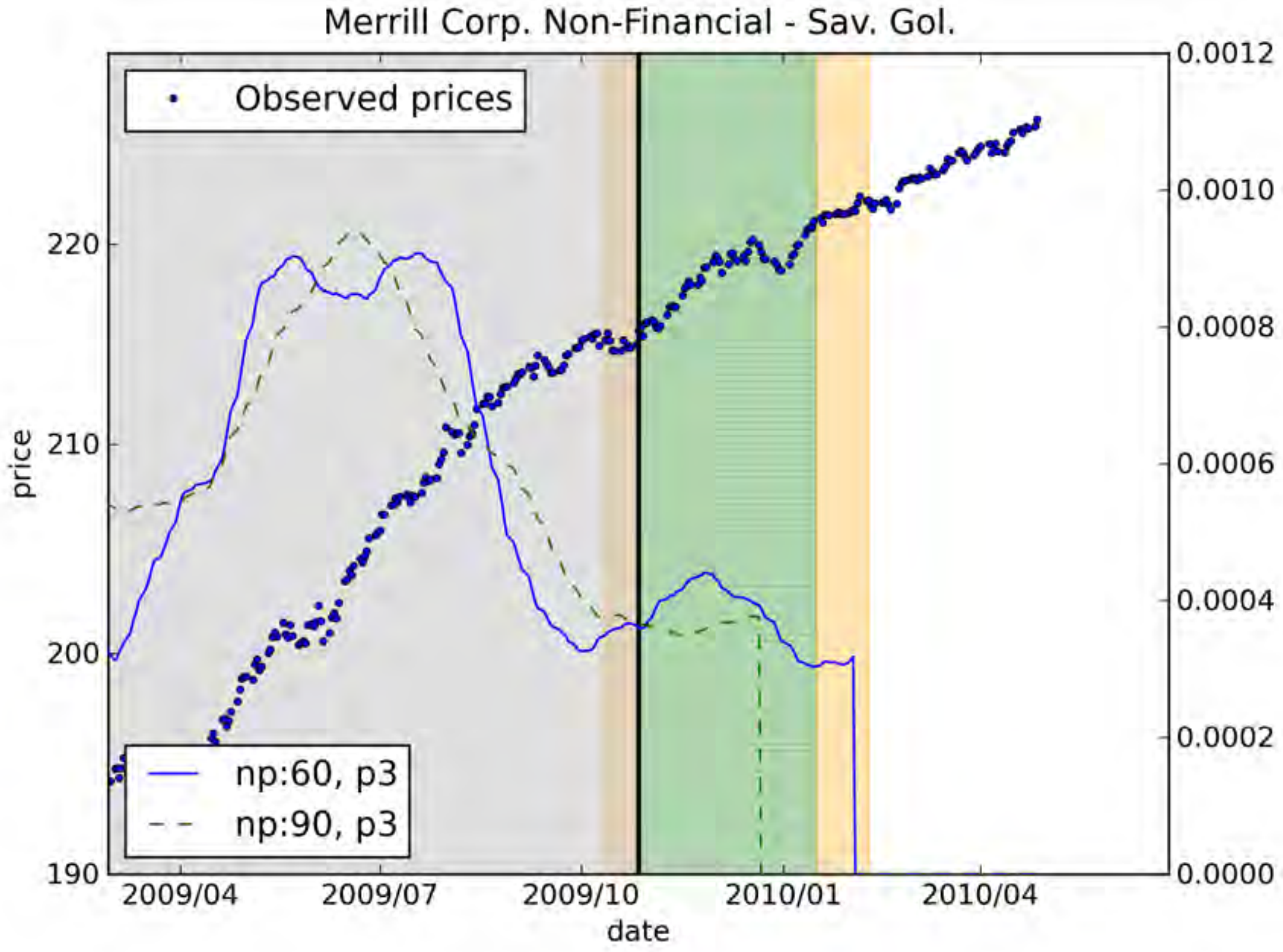

FIG. 7: On the data shown in figure 5, the dashed green line (respectively continuous blue line) give the smoothed growth rate of the Merrill Corp. Non-Financial index in the middle of a running window of 180 (respectively 120 days), as estimated with the Savitzky-Golay smoothing algorithm with a polynomial of order 3.

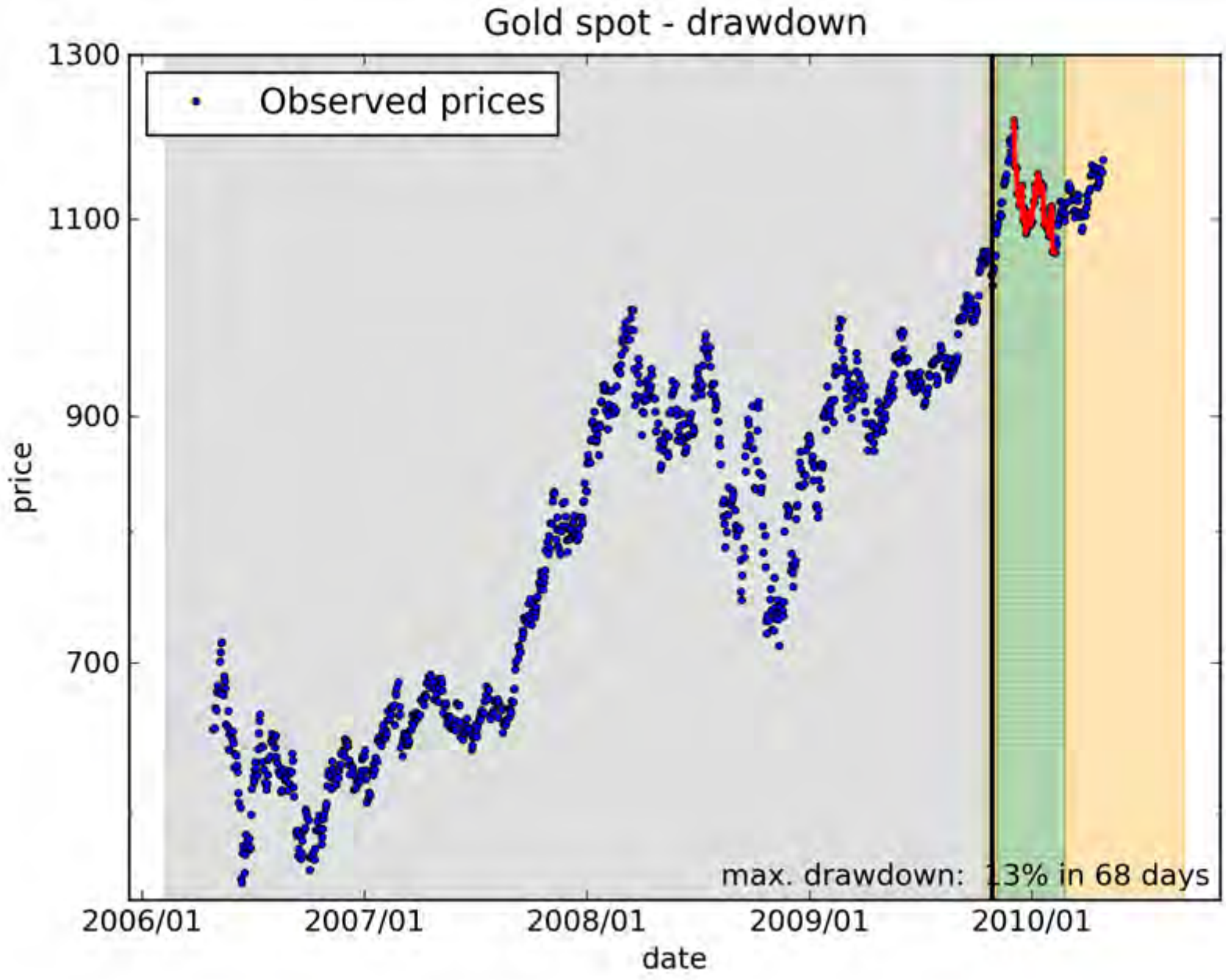

FIG. 8: Time series of gold spot price shown as filled blue circles, the 20-80% (respectively 5-95%) quantile intervals for the predicted end of the bubble and the subsequent evolution of the index. The black vertical line shows the time of the last observation used in the analysis. The red solid trace shows the largest drawdown that occurred after the forecast, 13% in 68 days.

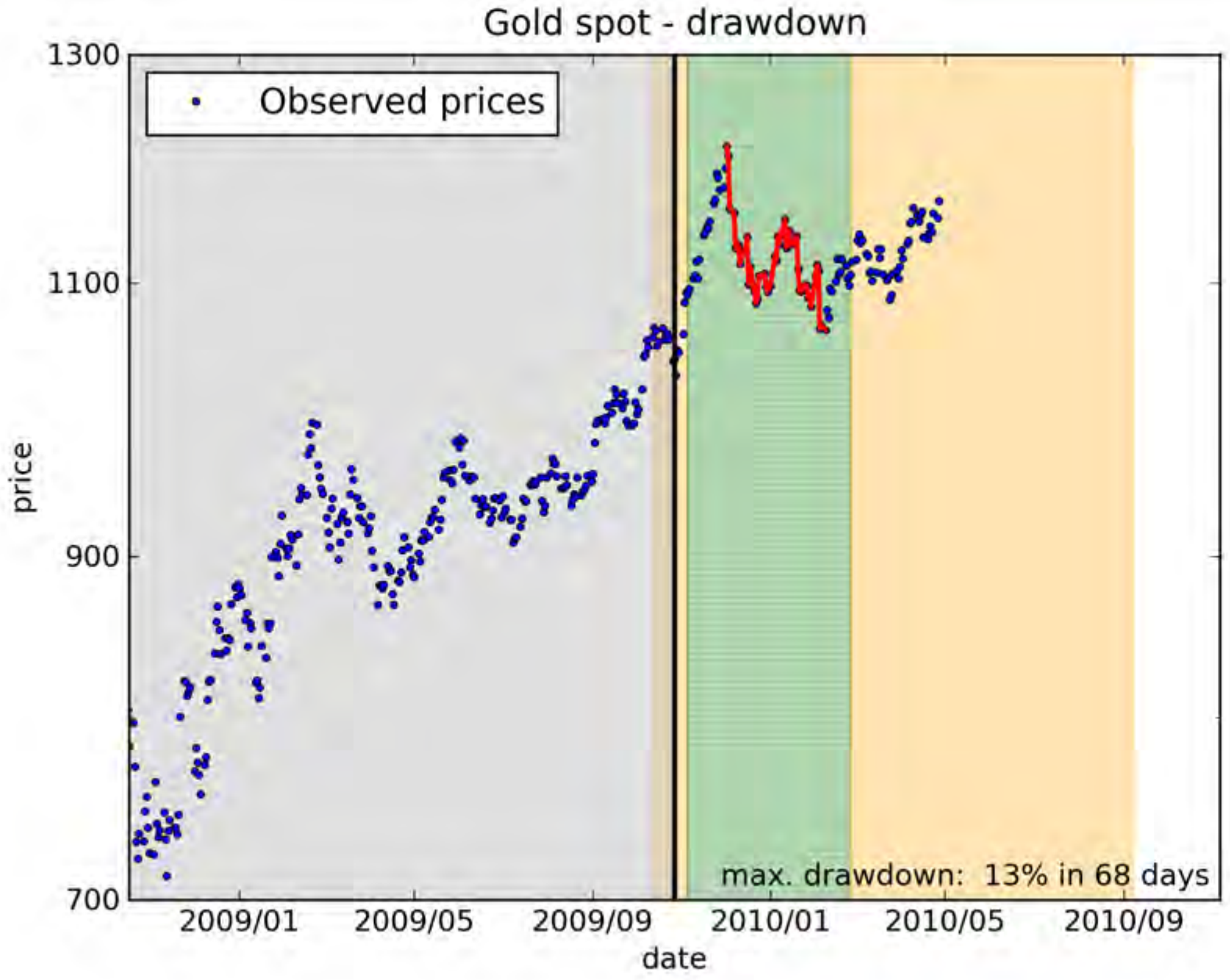

FIG. 9: Zoom of Figure 8.

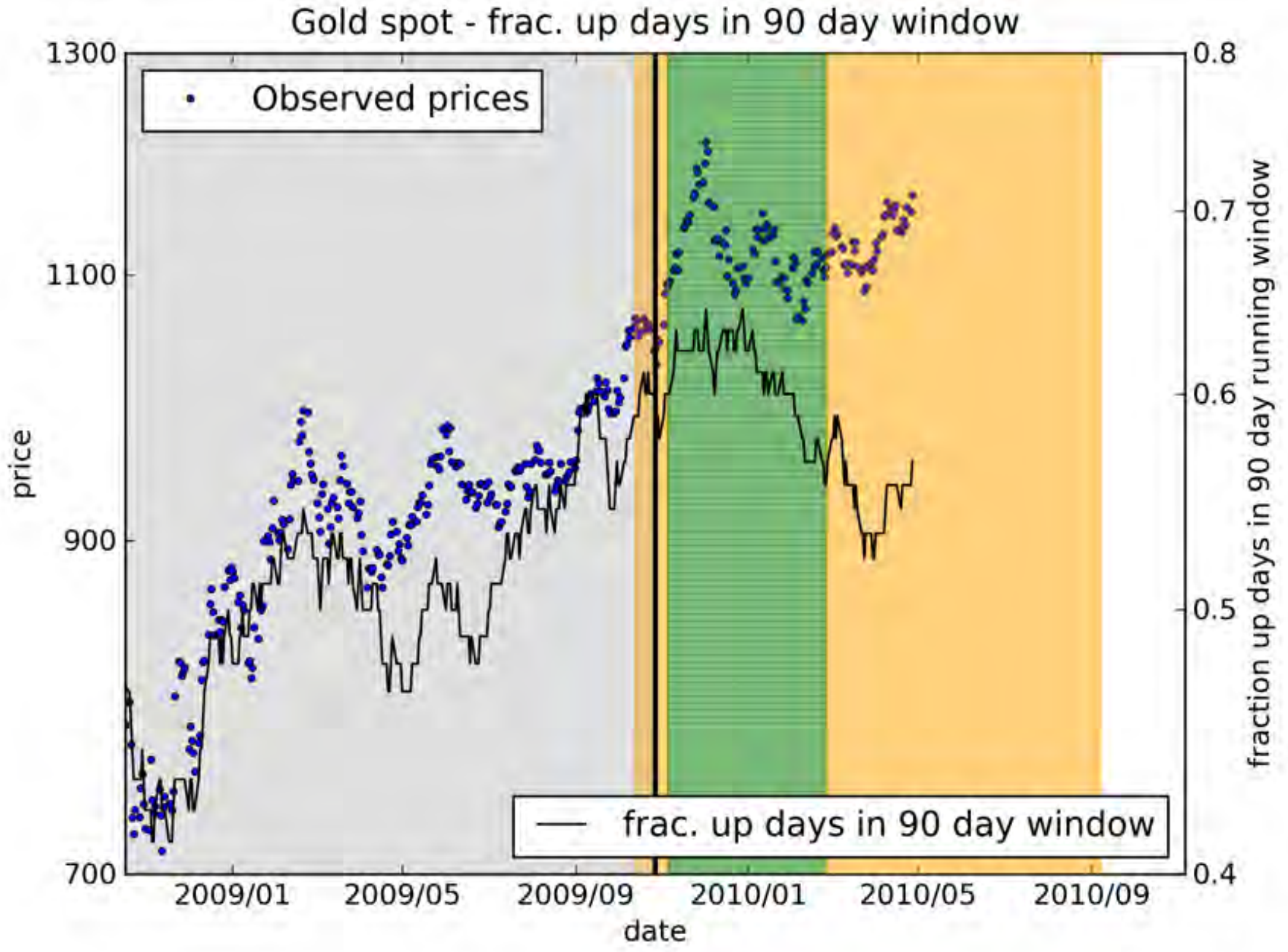

FIG. 10: On the data shown in figure 8, we plot the fraction of days (right vertical scale) with positive returns as a function of the right-end time of a moving window of width equal to 90 days.

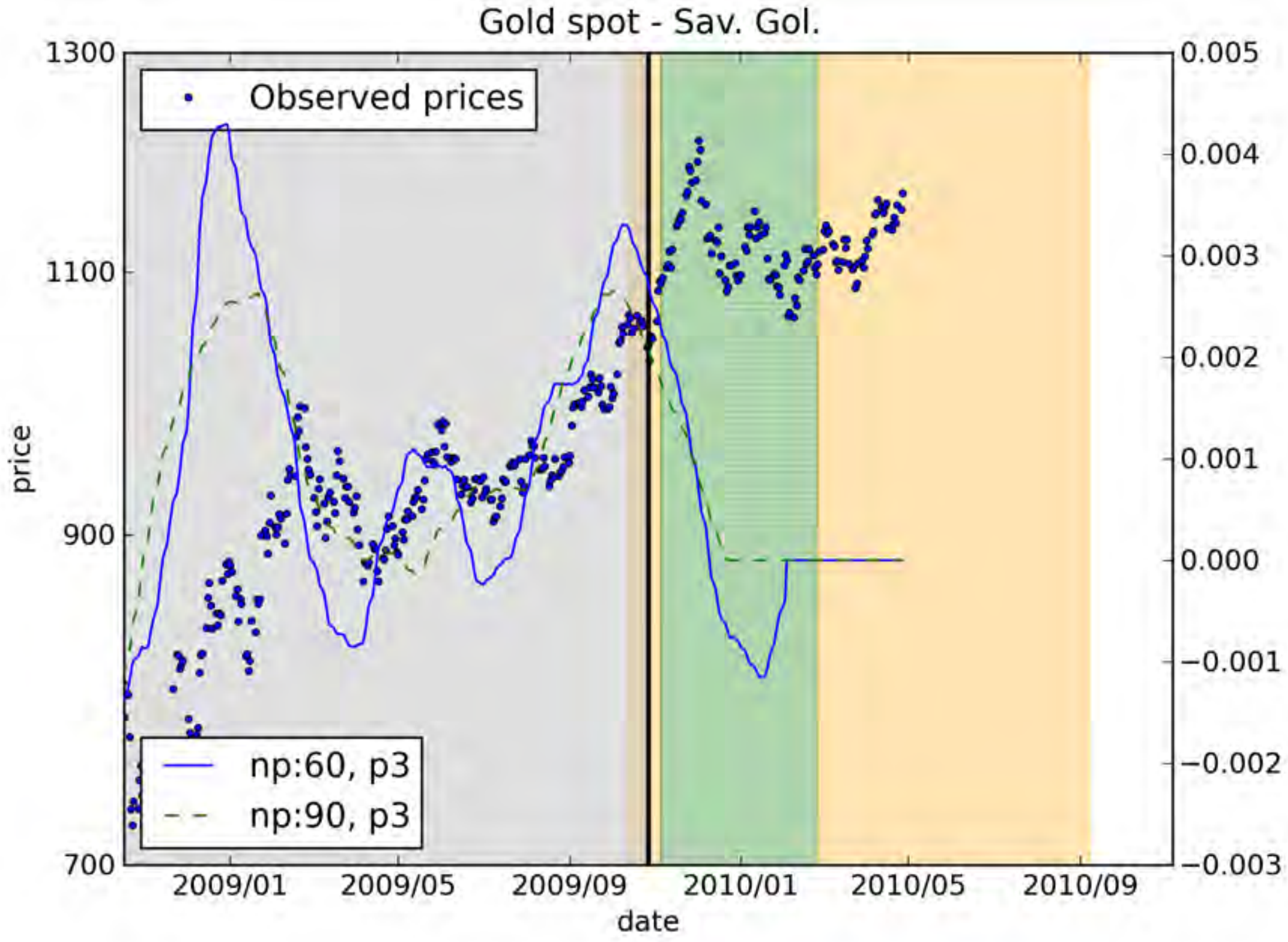

FIG. 11: On the data shown in Figure 8, the dashed green line (respectively continuous blue line) gives the smoothed growth rate of the gold spot price in the middle of a running window of 180 (respectively 120 days), as estimated with the Savitzky-Golay smoothing algorithm with a polynomial of order 3.

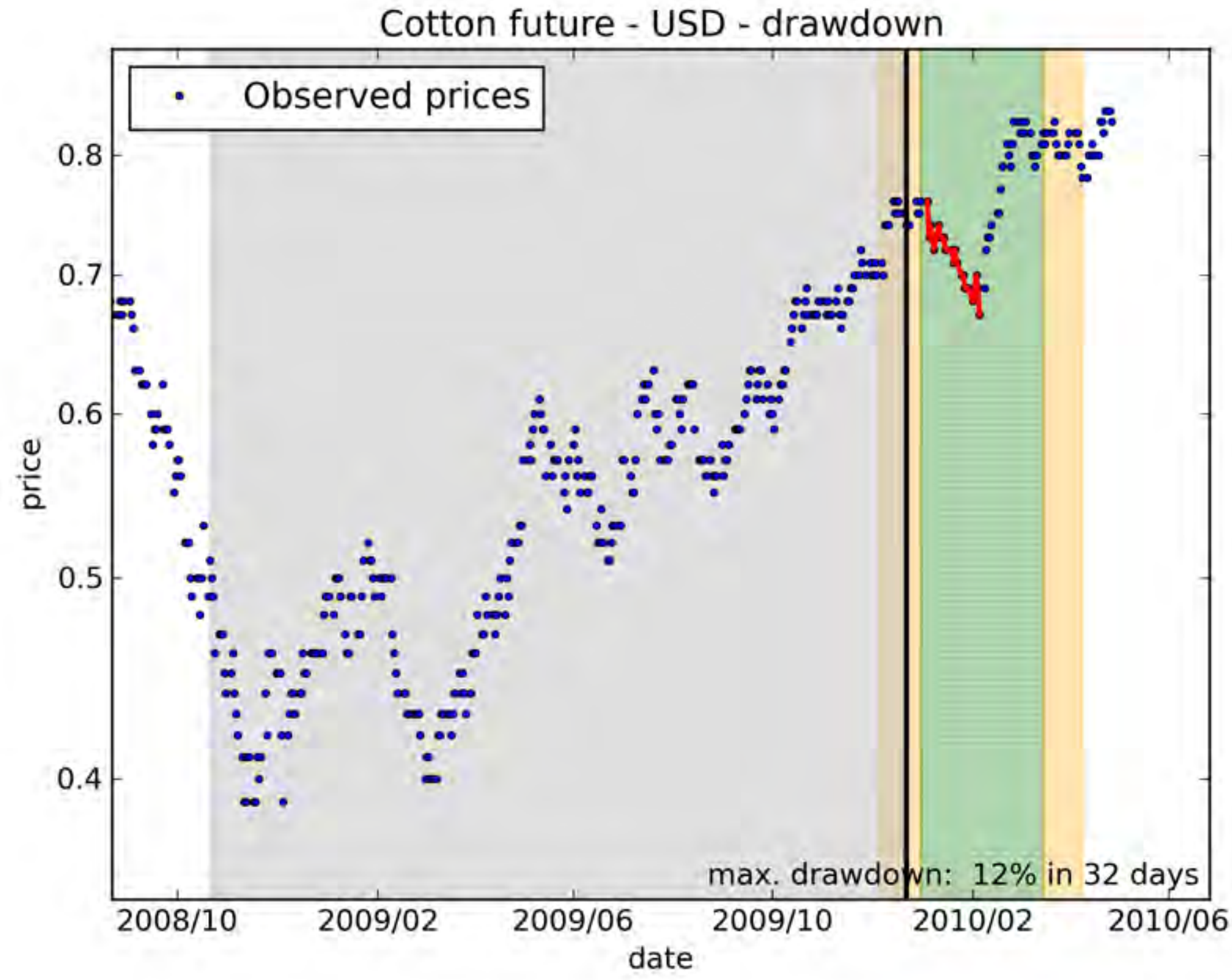

FIG. 12: Time series of the Cotton future price in USD shown as filled blue circles, the 20-80% (respectively 5-95%) quantile intervals for the predicted end of the bubble and the subsequent evolution of the index. The black vertical line shows the time of the last observation used in the analysis. The red solid trace shows the largest drawdown that occurred after the forecast, 12% in 32 days.

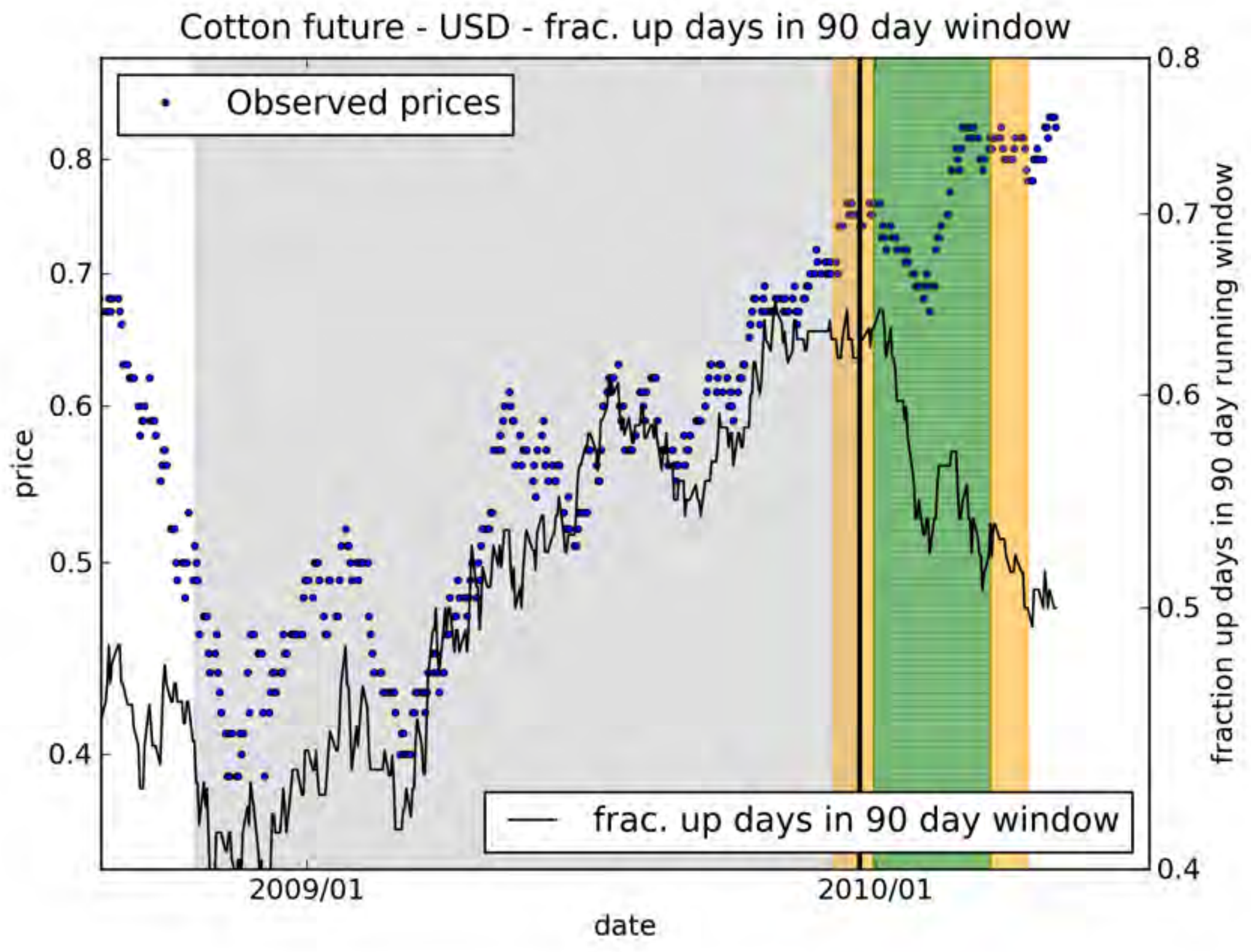

FIG. 13: On the data shown in figure 12, we plot the fraction of days (right vertical scale) with positive returns as a function of the right-end time of a moving window of width equal to 90 days.

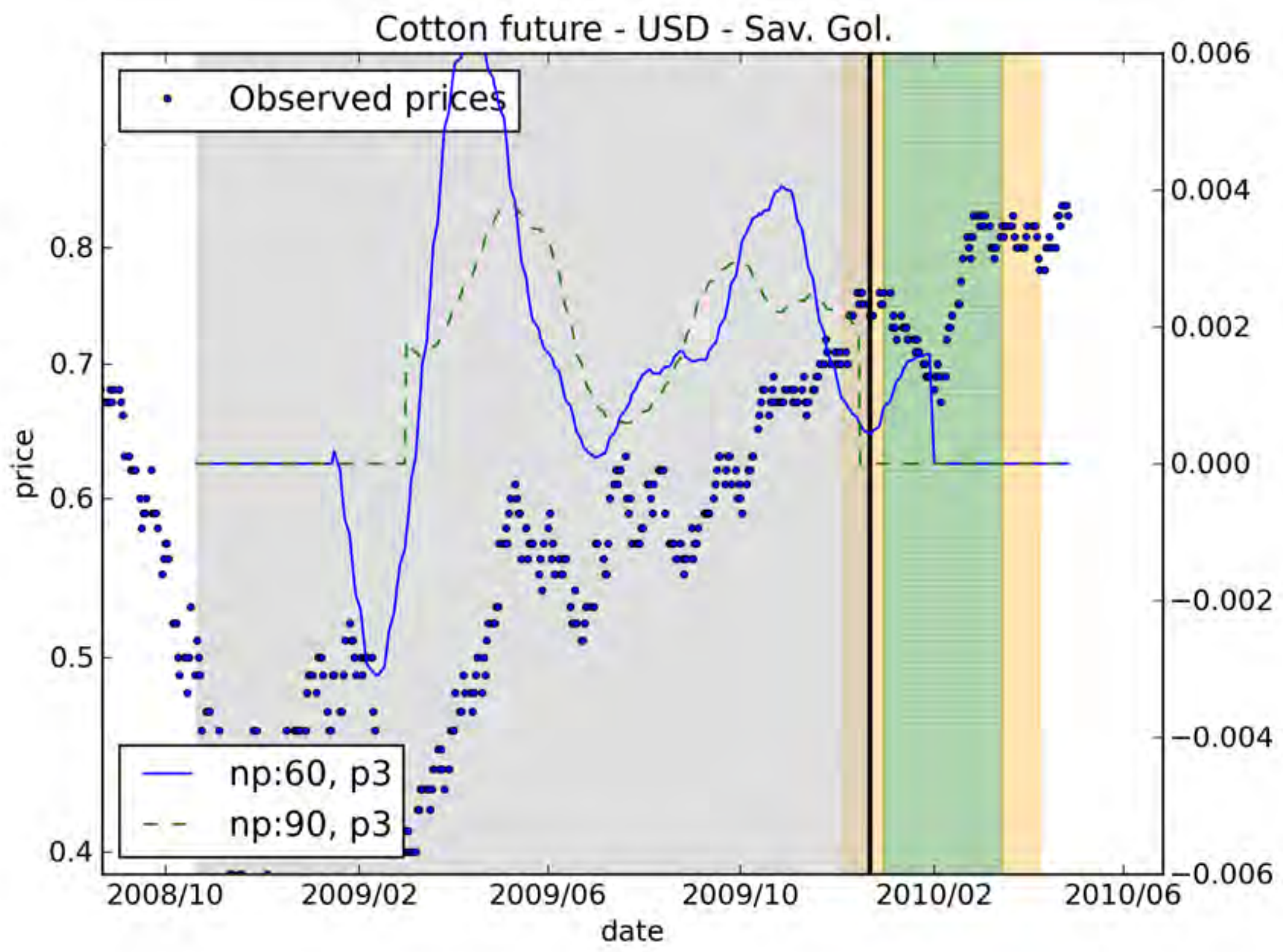

FIG. 14: On the data shown in Figure 12, the dashed green line (respectively continuous blue line) gives the smoothed growth rate of the cotton futures in the middle of a running window of 180 (respectively 120 days), as estimated with the Savitzky-Golay smoothing algorithm with a polynomial of order 3.

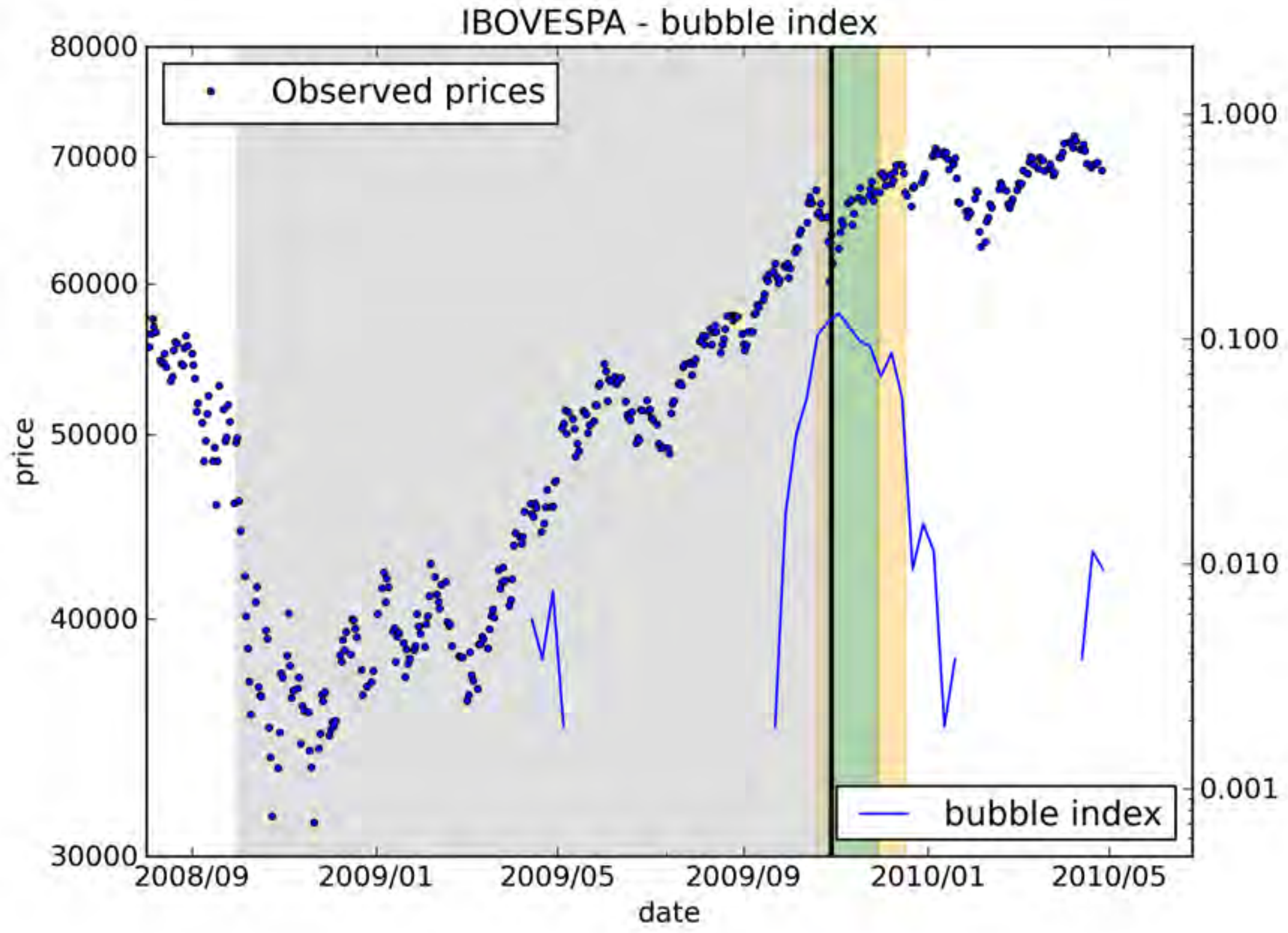

FIG. 15: On the data shown in Figure 1, our recently derived “bubble index”.

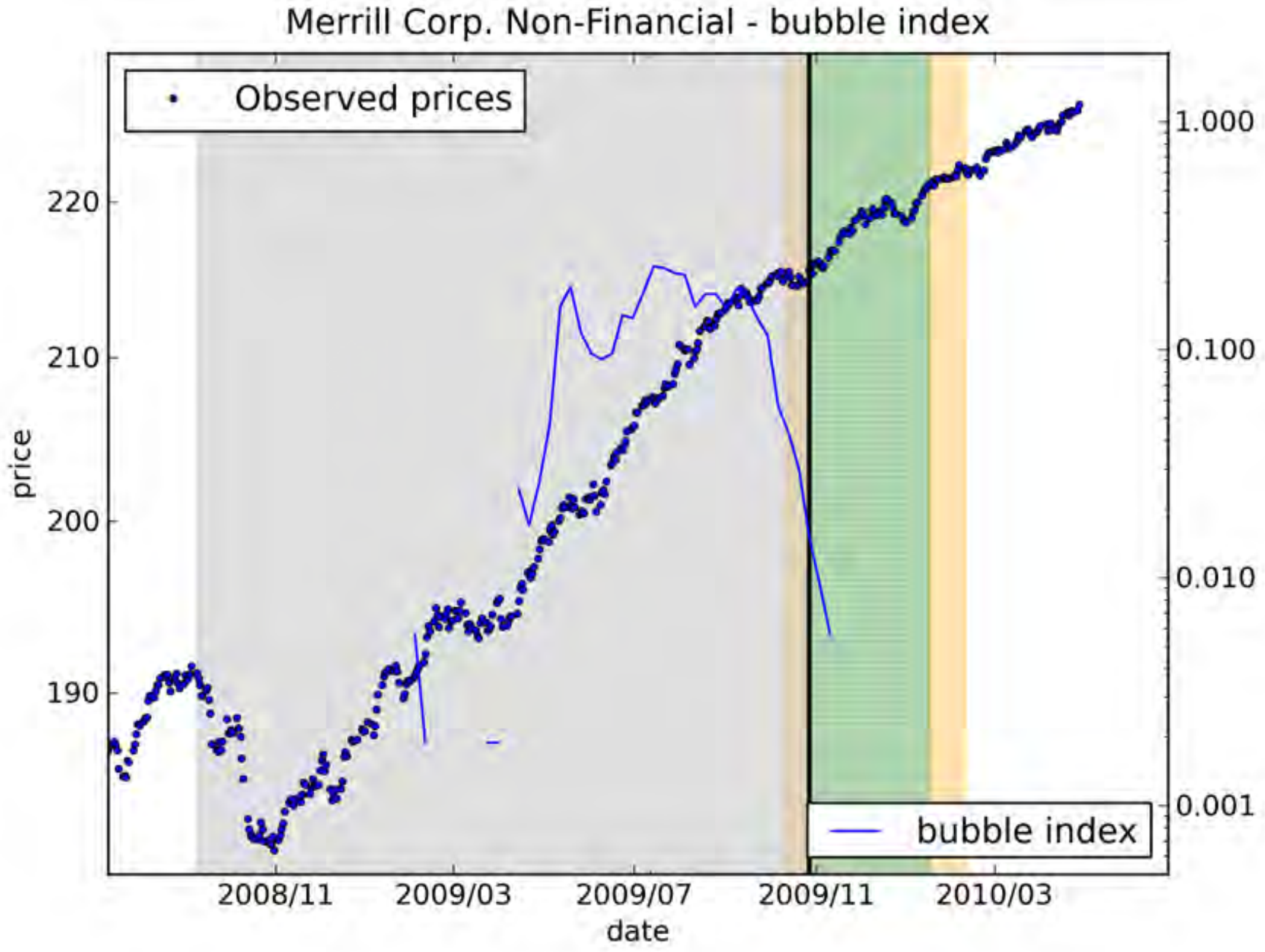

FIG. 16: On the data shown in Figure 5, our recently derived “bubble index”.

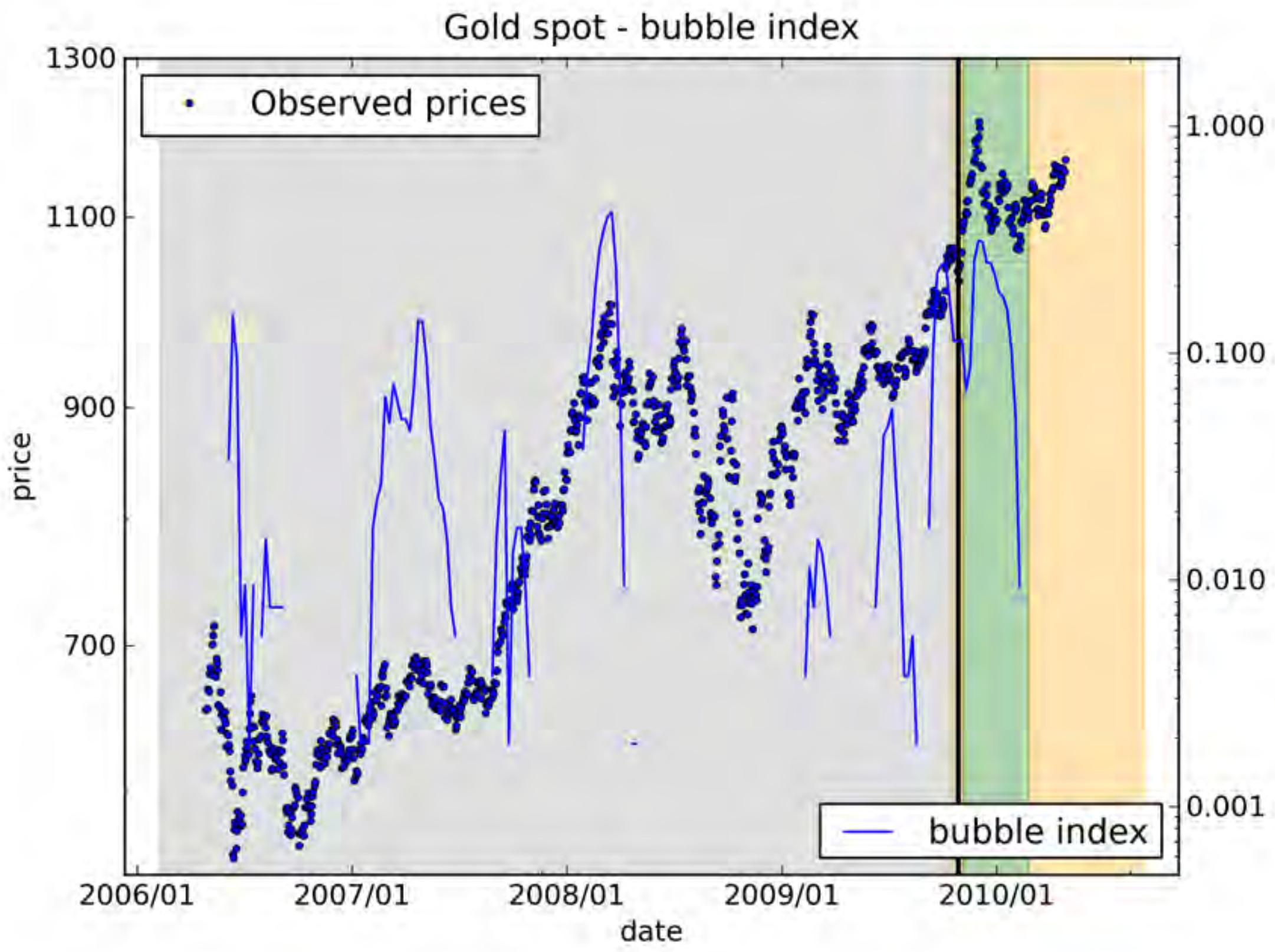

FIG. 17: On the data shown in Figure 8, our recently derived “bubble index”.

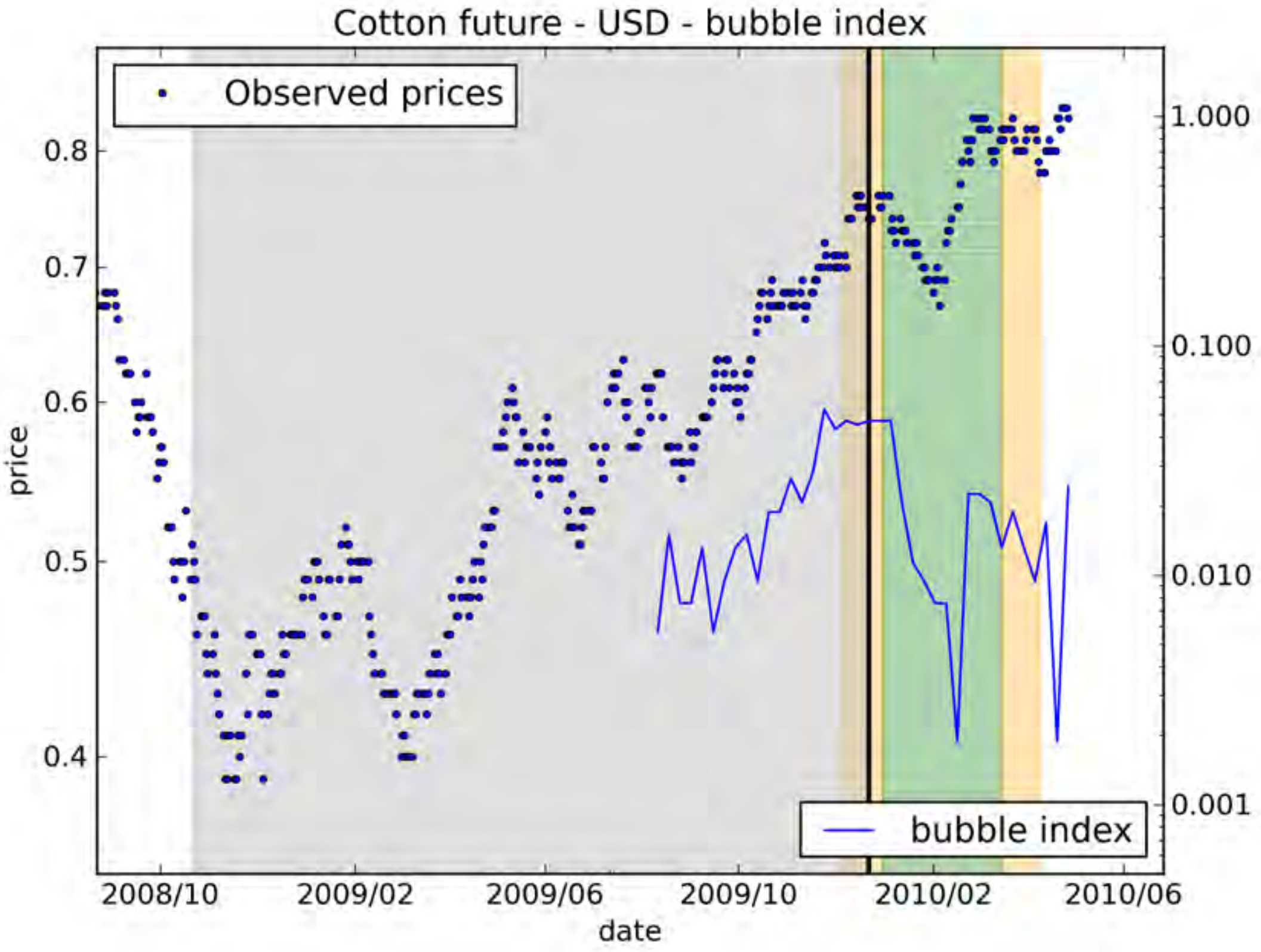

FIG. 18: On the data shown in Figure 12, our recently derived "bubble index".

# FCO Report: Brazil Index (BVSP)

Financial Crisis Observatory, ETH Zurich

November 02, 2009

## 1 Data source

We performed our analysis on the IBOVESPA index on the Sao Paolo, Brazil exchange. We acquired the data from http://finance.yahoo.com/ using the ticker '^BVSP' and the date range 2008-09-01 through 2009-10-29.

## 2 Input Parameters

| | |
|---|---|
| Date of last observation used in analysis | 2009-10-29 |
| Date of observed peak of data | 2008-05-20 |
| Number LPPL intervals found | 69 |
| Number total intervals tested | 212 |

## 3 Forecast quantiles for $t_c$

| | **Low** | **High** |
|---|---|---|
| 05/95 | 2009-10-19 | 2009-12-17 |
| 20/80 | 2009-10-27 | 2009-11-29 |

## 4 Plots of observations, fits and forecasts

Guide to figures:

- Observations appear as filled circles.
- Shaded regions:
  - Lightest, hashed with circles: region of $t_2$ used in analysis
  - Mid-grey, hashed with horizontal lines: region of 5%/95% forecast quantiles of $t_c$.
  - Darkest, hashed with diagonal lines: region of 20%/80% forecast quantiles of $t_c$.
- Lines:
  - Solid lines before final observation are LPPL fits to observations.
  - Lines after final observation are extrapolations. They are dashed lines except for within 20 days on either side of fit parameter $t_c$.

Successive figures are zoomed-in versions of previous figures.

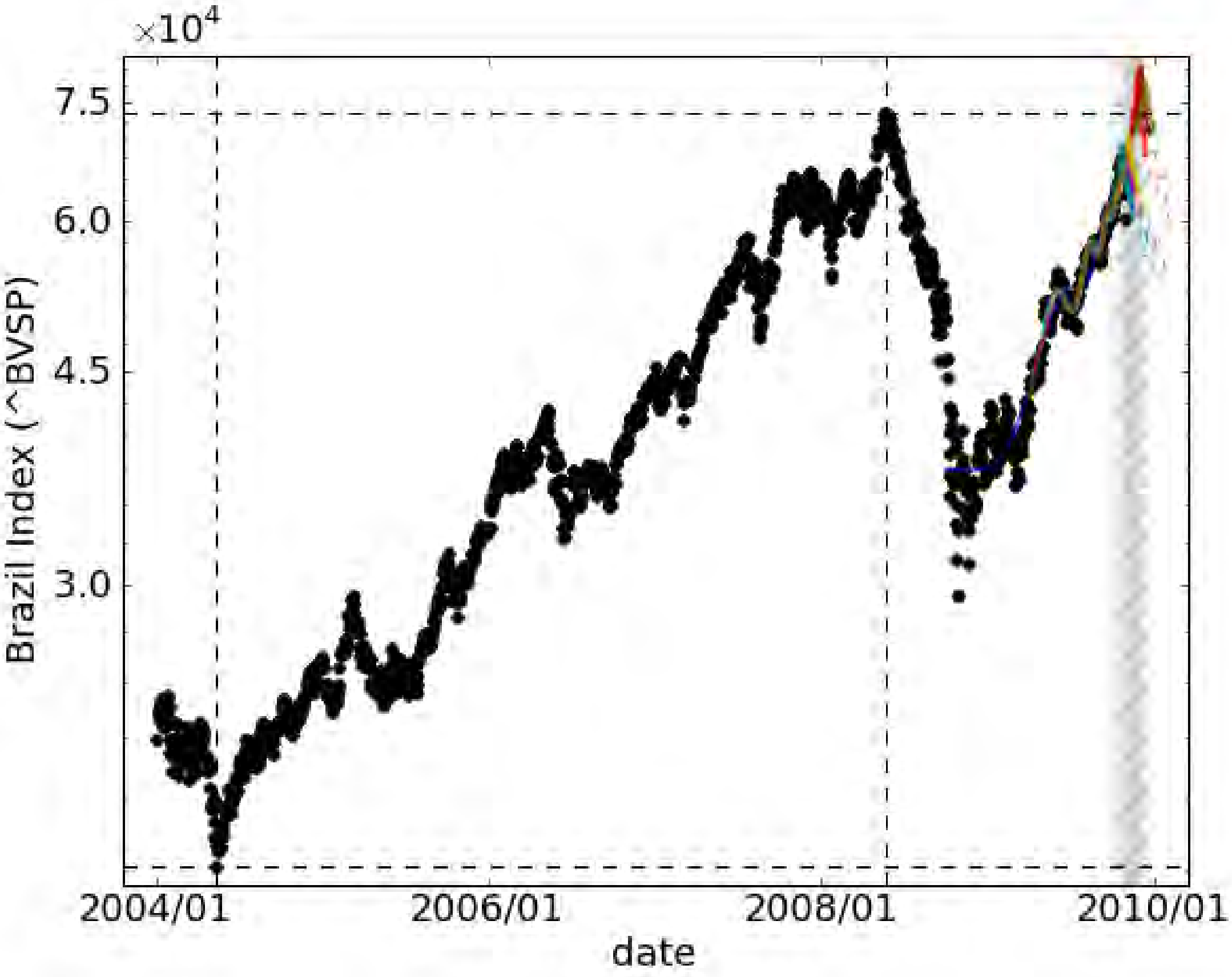

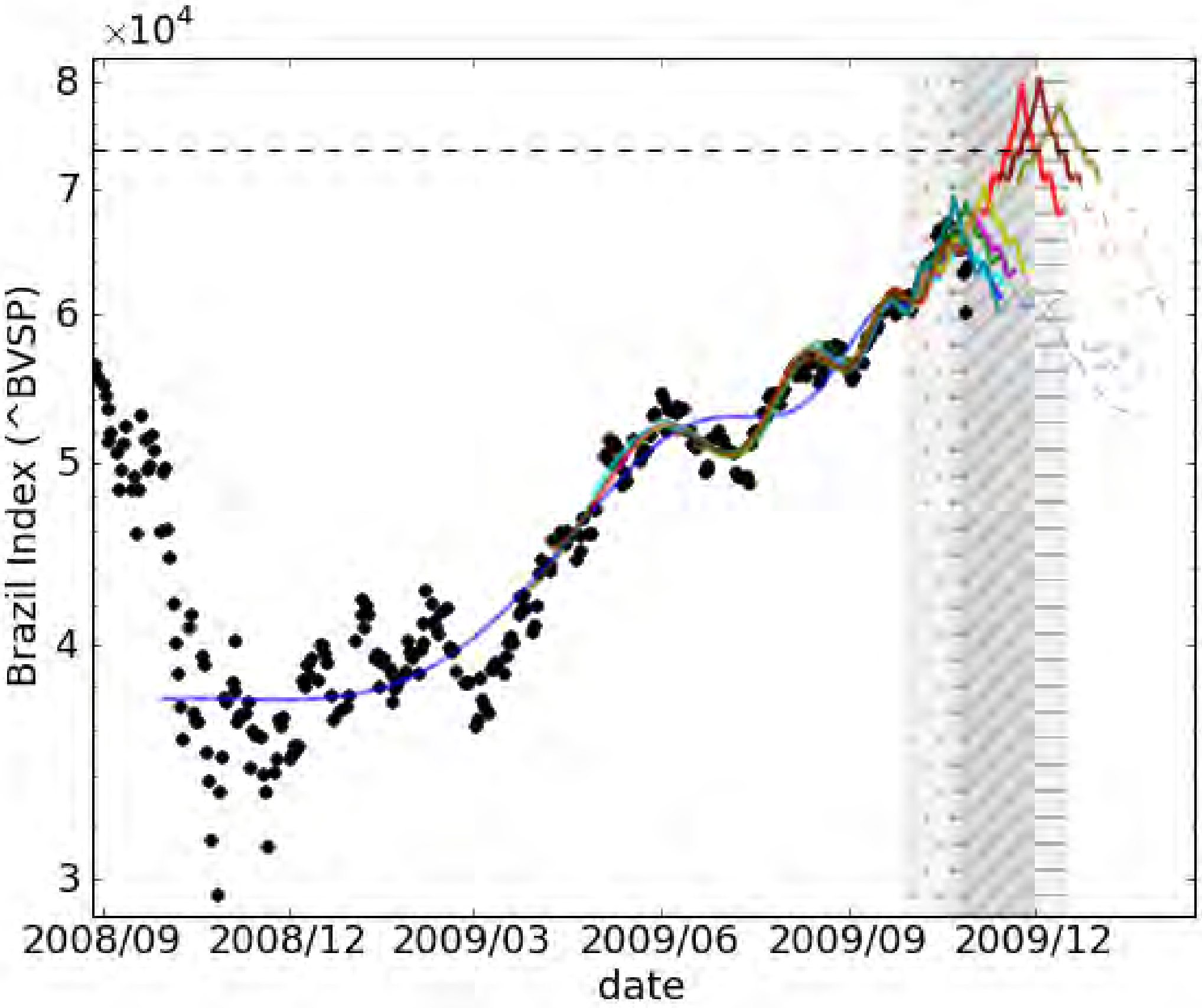

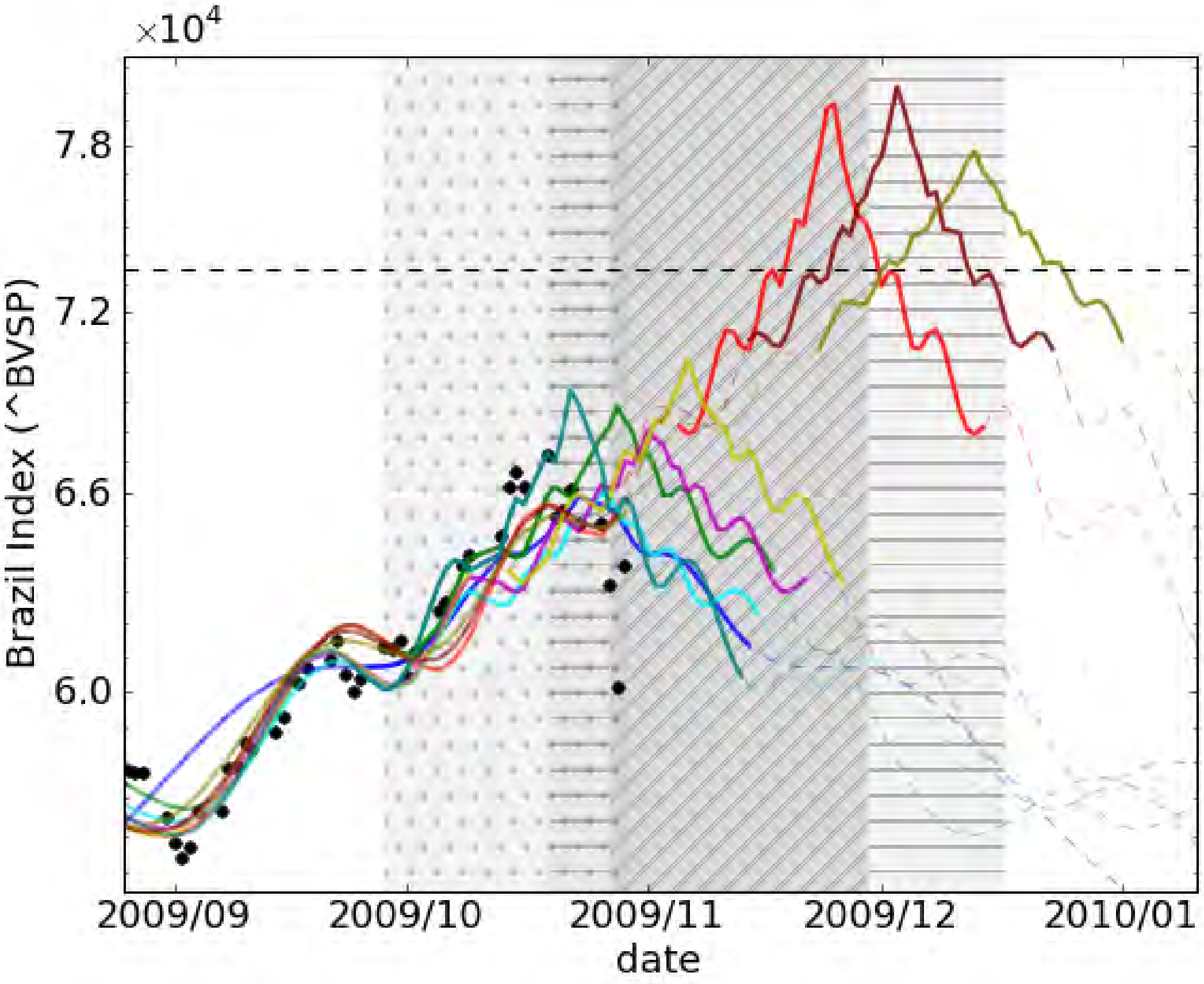

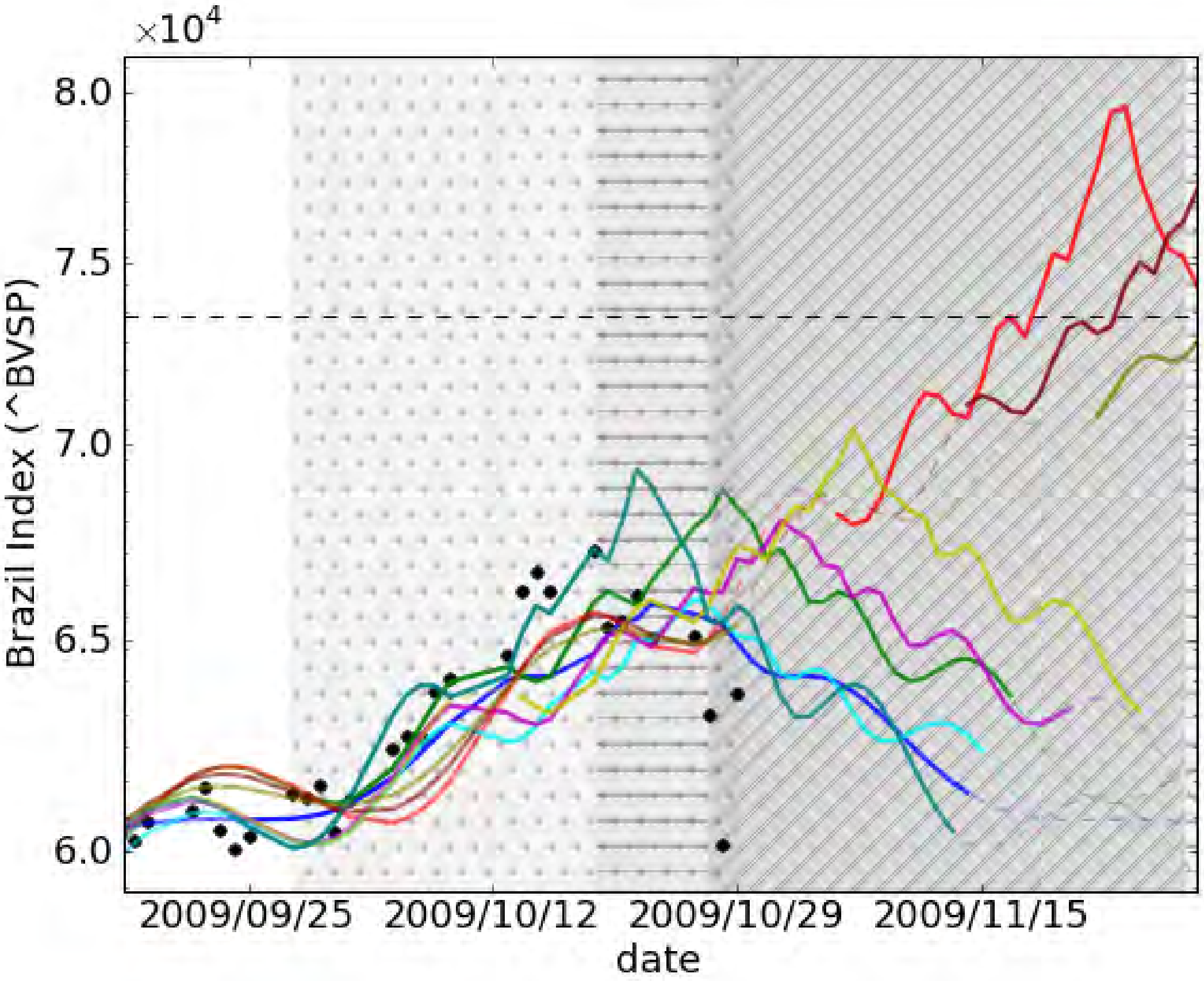

# FCO Report: Merrill Lynch Corp. Non-Financial Bond Index

Financial Crisis Observatory, ETH Zurich

November 02, 2009

## 1 Data source

We performed our analysis on the Total Return Index column of the Merrill Lynch "EMU Corporates, Non-Financial Index" (where EMU is European Monetary Union). We downloaded the data from a Bloomberg terminal, where the code for this Index is 'EN00'. We analyzed data from 2007-01-01 through 2009-10-27, though only the data after late 2008 is relevant to the results.

## 2 Input Parameters

| | |
|---|---|
| Date of last observation used in analysis | 2009-10-27 |
| Date of observed peak of data | 2009-10-27 |
| Number LPPL intervals found | 11 |
| Number total intervals tested | 619 |

## 3 Forecast quantiles for $t_c$

| | **Low** | **High** |
|---|---|---|
| 05/95 | 2009-10-11 | 2010-02-09 |
| 20/80 | 2009-10-27 | 2010-01-16 |

## 4 Plots of observations, fits and forecasts

Guide to figures:

- Observations appear as filled circles.
- Shaded regions:
  - Lightest, hashed with circles: region of $t_2$ used in analysis
  - Mid-gray, hashed with horizontal lines: region of 5%/95% forecast quantiles of $t_c$.
  - Darkest, hashed with diagonal lines: region of 20%/80% forecast quantiles of $t_c$.
- Lines:
  - Solid lines before final observation are LPPL fits to observations.
  - Lines after final observation are extrapolations. They are dashed lines except for within 20 days on either side of fit parameter $t_c$.

Successive figures are zoomed-in versions of previous figures.

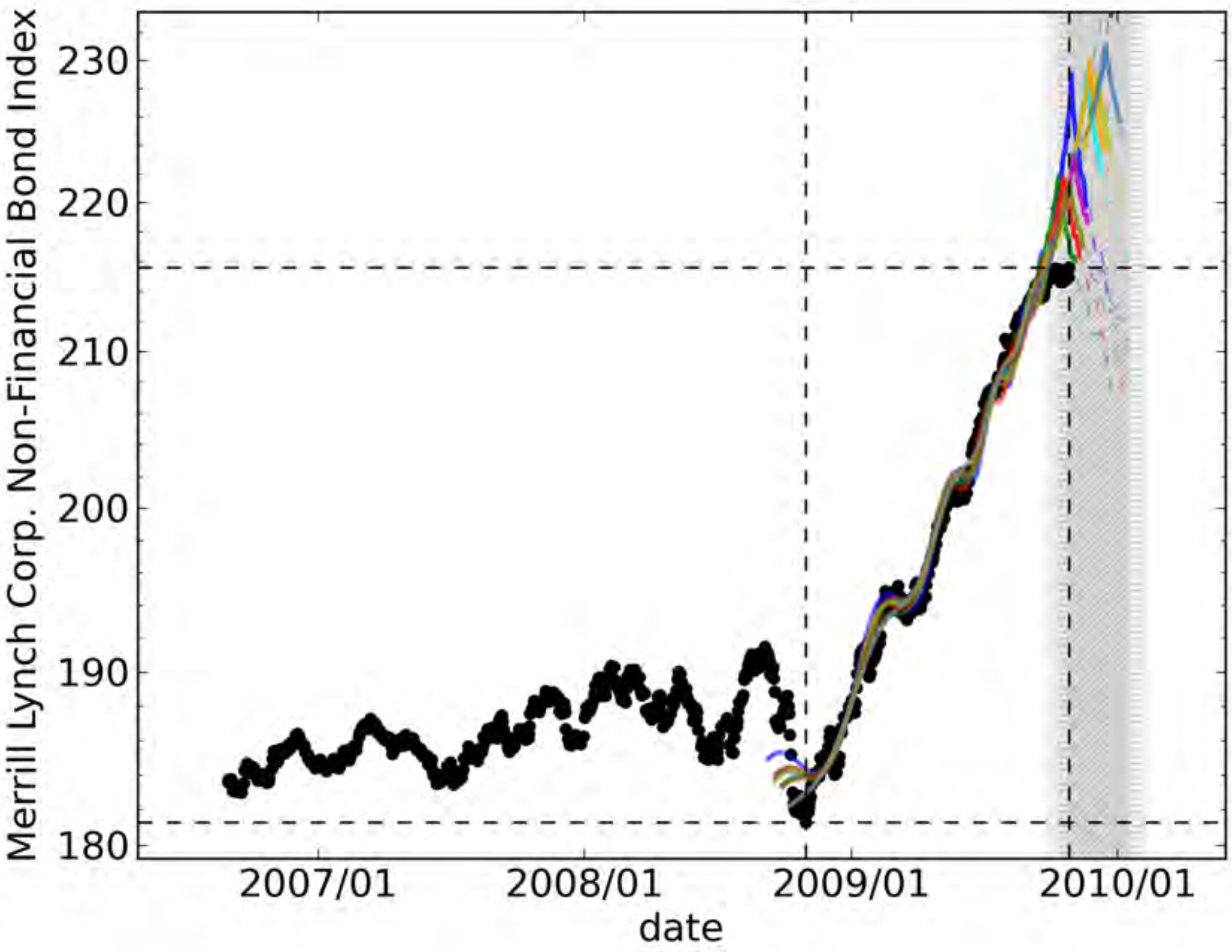

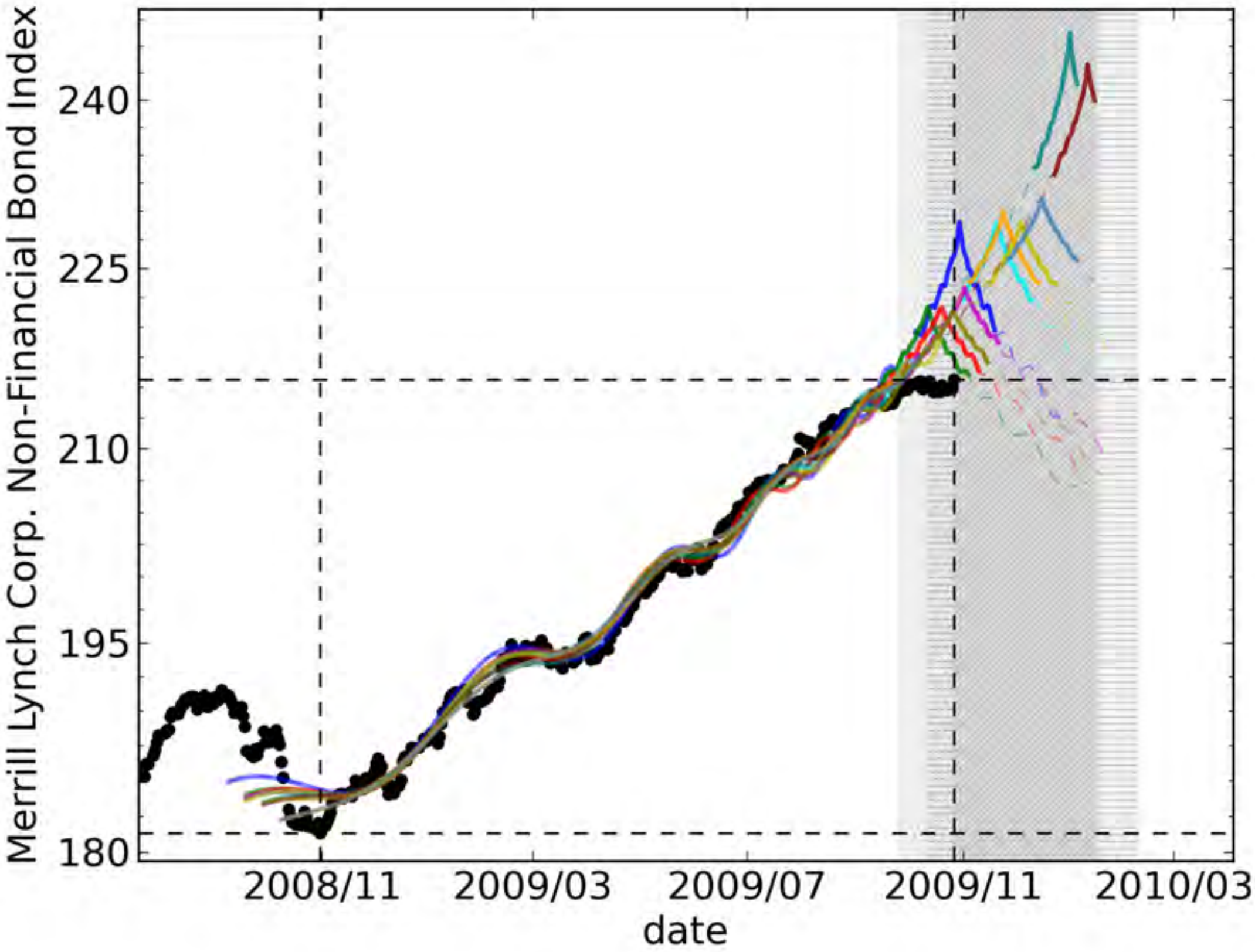

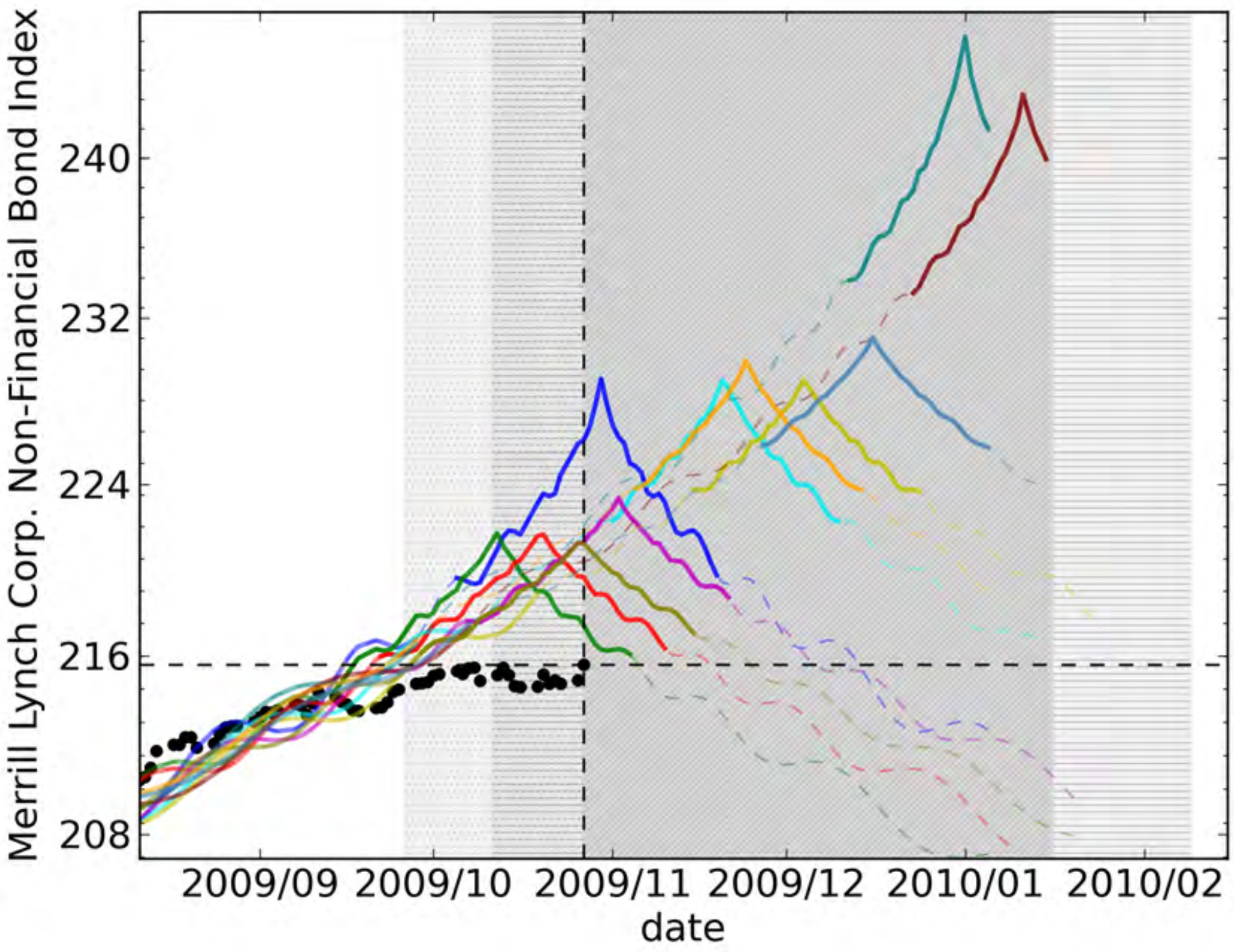

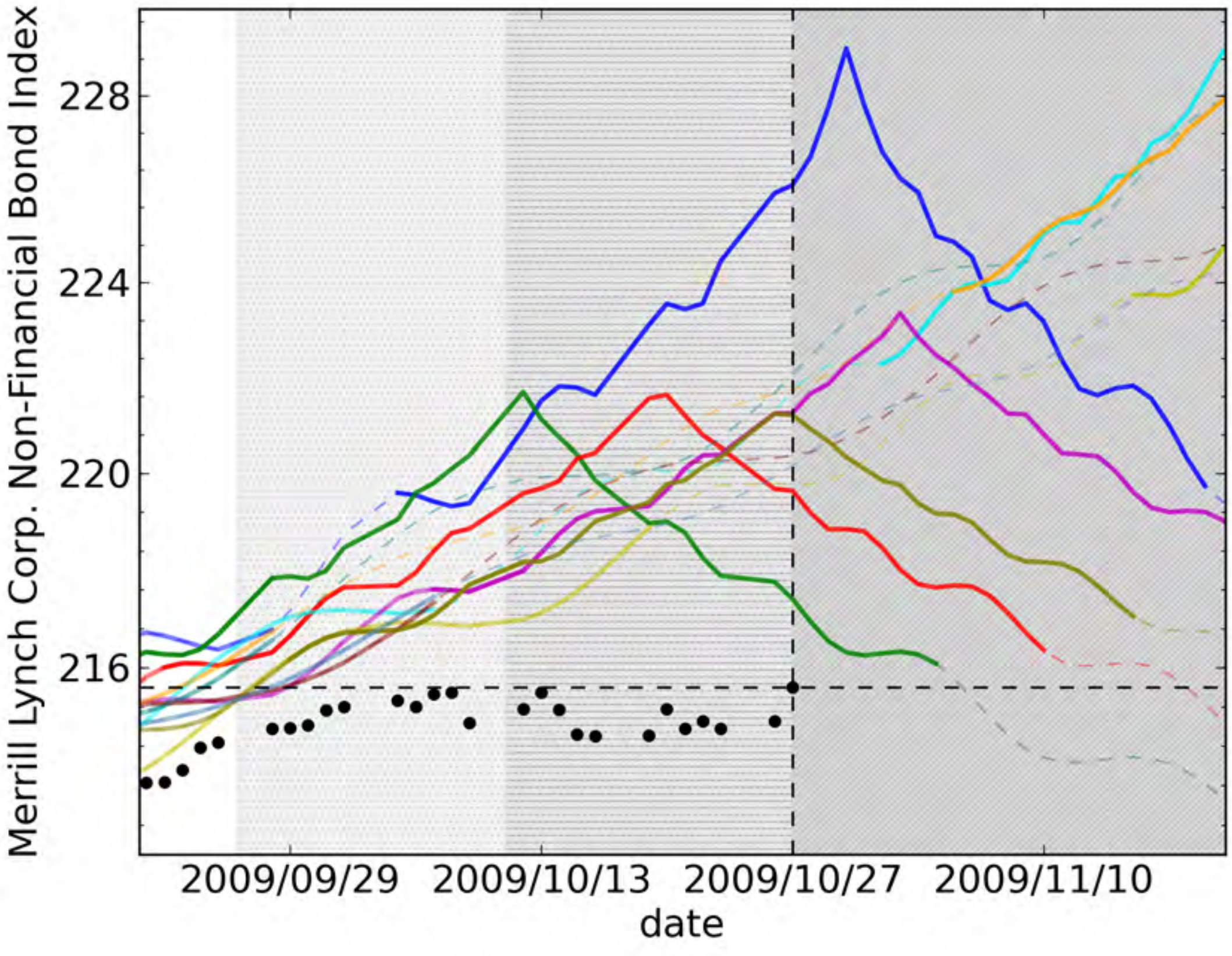

# FCO Report: Gold spot price $/oz

Financial Crisis Observatory, ETH Zurich

November 02, 2009

## 1 Data source

We performed our analysis on the spot price of gold in USD/oz. We acquired the time series from a Bloomberg terminal for the range 2006-01-01 through 2009-10-27.

## 2 Input Parameters

| | |
|---|---|
| Date of last observation used in analysis | 2009-10-27 |
| Date of observed peak of data | 2009-10-13 |
| Number LPPL intervals found | 118 |
| Number total intervals tested | 417 |

## 3 Forecast quantiles

| | **Low** | **High** |
|---|---|---|
| 05/95 | 2009-10-13 | 2010-09-07 |
| 20/80 | 2009-11-05 | 2010-02-25 |

## 4 Plots of observations, fits and forecasts

Guide to figures:

- Observations appear as filled circles.
- Shaded regions:
  - Lightest, hashed with circles: region of $t_2$ used in analysis
  - Mid-gray, hashed with horizontal lines: region of 5%/95% forecast quantiles of $t_c$.
  - Darkest, hashed with diagonal lines: region of 20%/80% forecast quantiles of $t_c$.
- Lines:
  - Solid lines before final observation are LPPL fits to observations.
  - Lines after final observation are extrapolations. They are dashed lines except for within 20 days on either side of fit parameter $t_c$.

Successive figures are zoomed-in versions of previous figures.

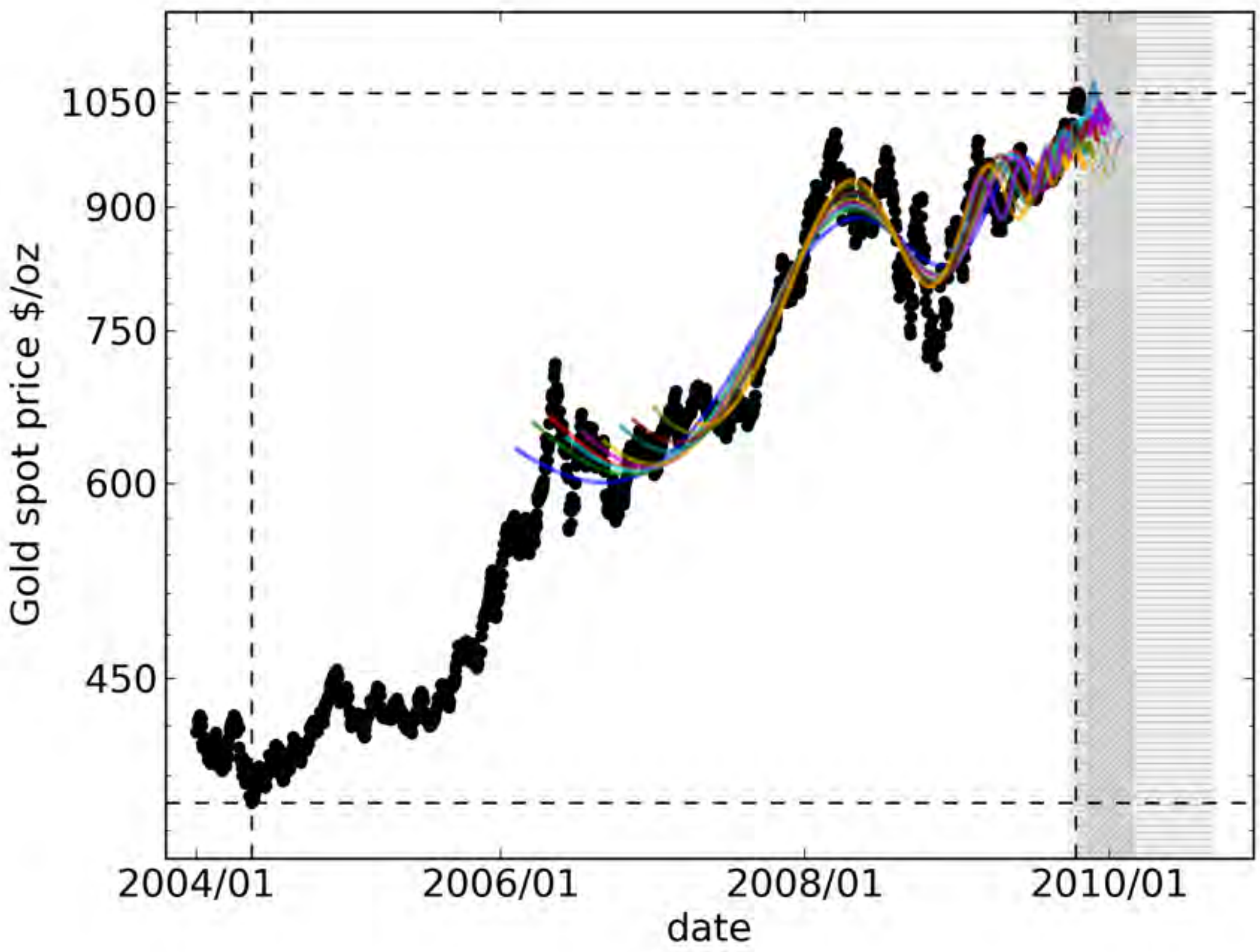

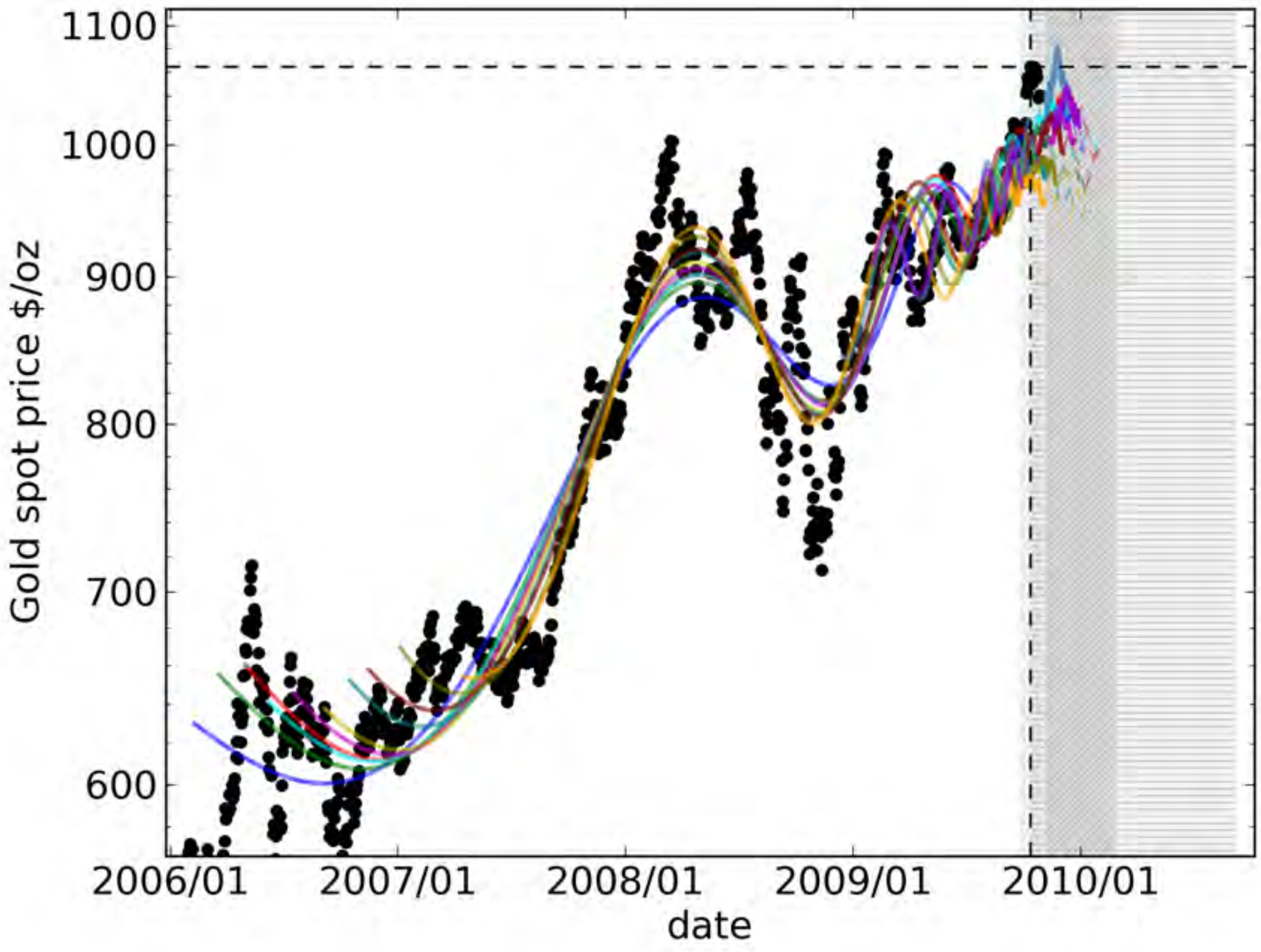

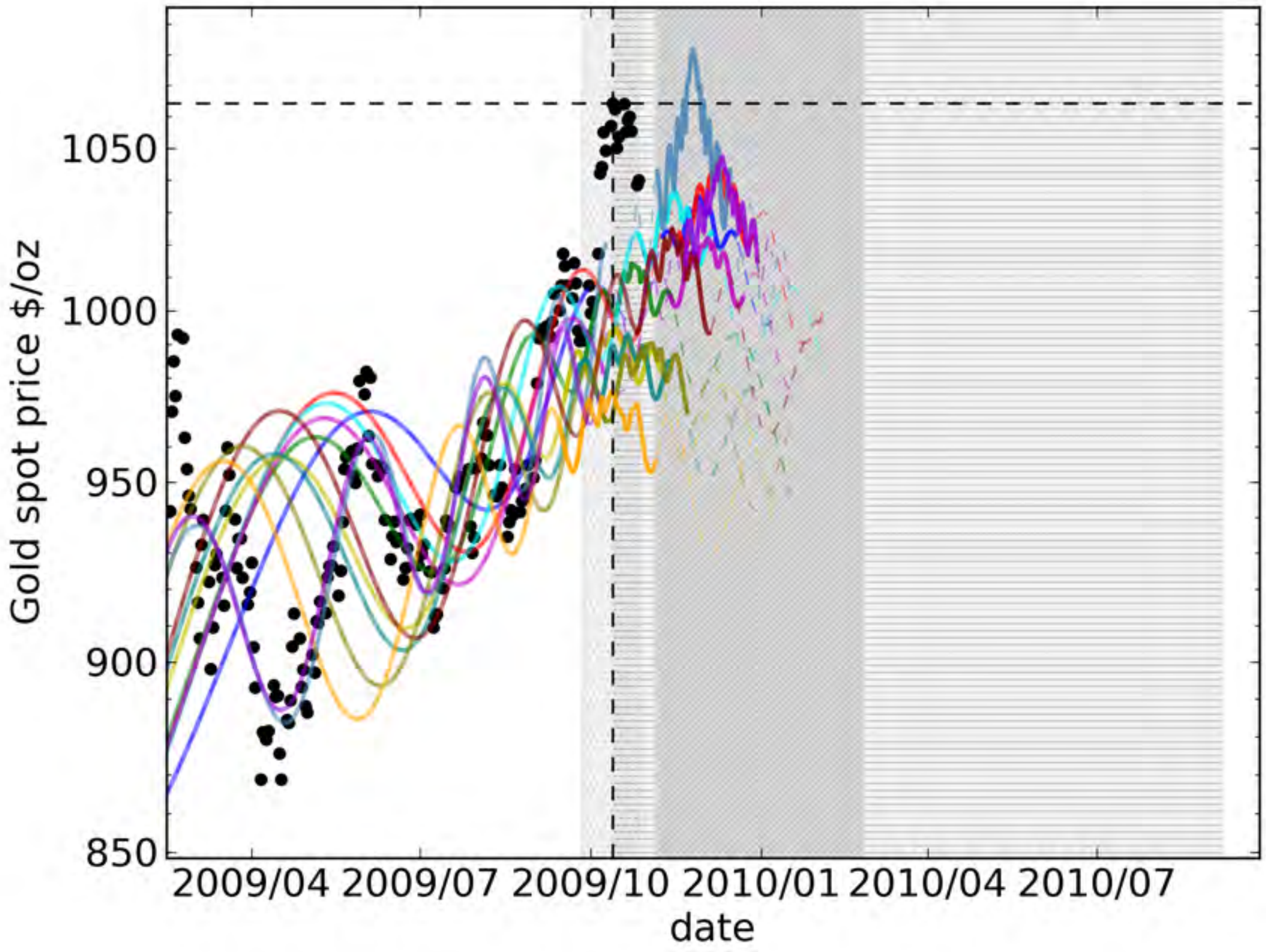

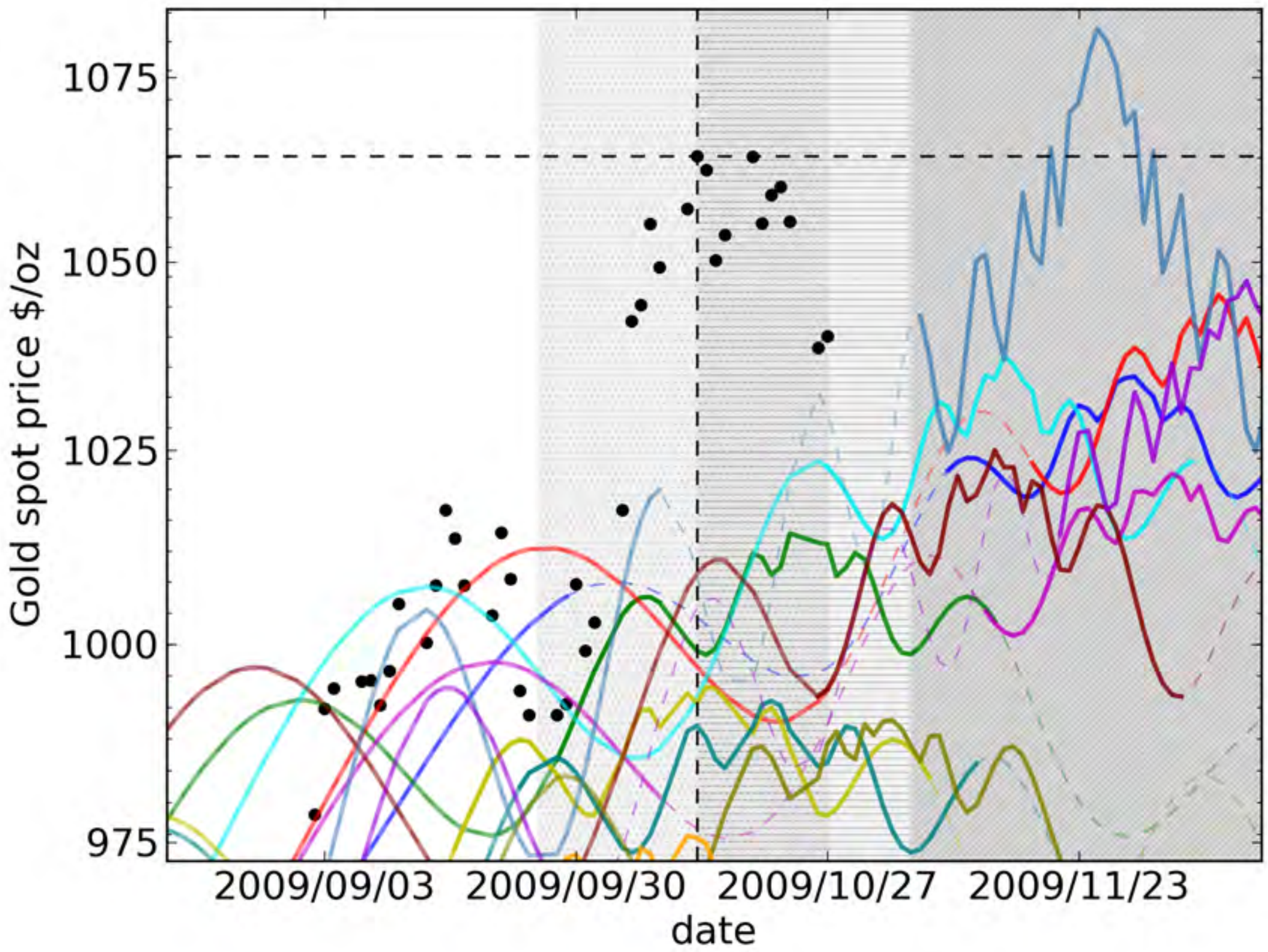

Financial Crisis Observatory
FCO @ ETH Zurich
2009-12-22

# 1 LPPL Analysis of Cotton, CT1 Comdty, Generic 1st ‘CT’ Future

## 1.1 Forecast quantiles

| | Low | High |
|---|---|---|
| 05/95 | 2009-12-05 | 2010-04-09 |
| 20/80 | 2009-12-31 | 2010-03-16 |

## 1.2 Plots of observations, fits and forecasts

Guide to figures:

- Observations appear as filled circles.
- Shaded regions:
- Lightest, hashed with circles: region of $t_2$ used in analysis
- Mid-grey, hashed with horizontal lines: region of 5%/95% forecast quantiles of $t_c$.
- Darkest, hashed with diagonal lines: region of 20%/80% forecast quantiles of $t_c$.
- Lines:
- Solid lines before final observation are LPPL fits to observations.
- Lines after final observation are extrapolations. They are dashed lines except for within 20 days on either side of fit parameter $t_c$.

Successive figures are zoomed-in versions of previous figures.

v0

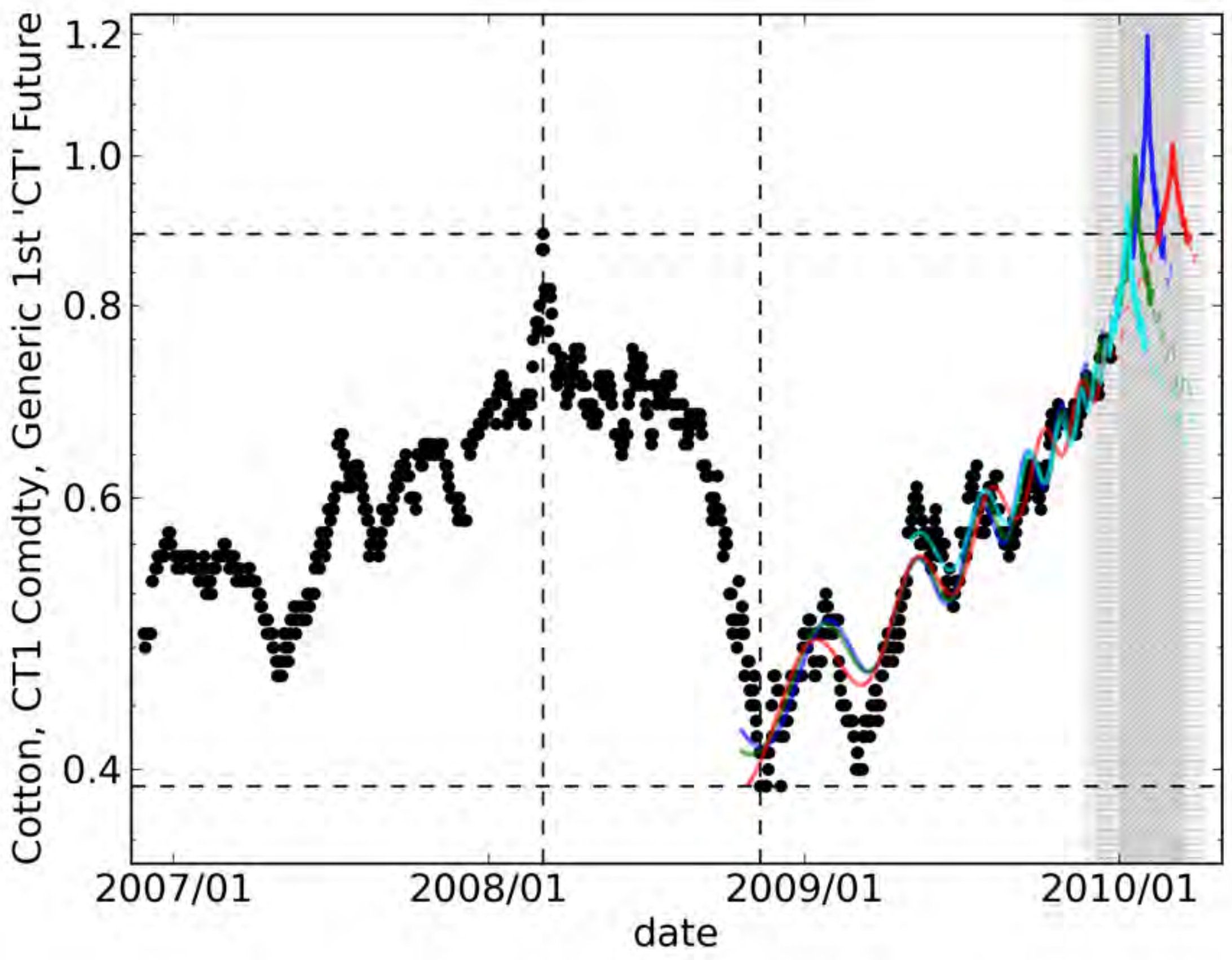

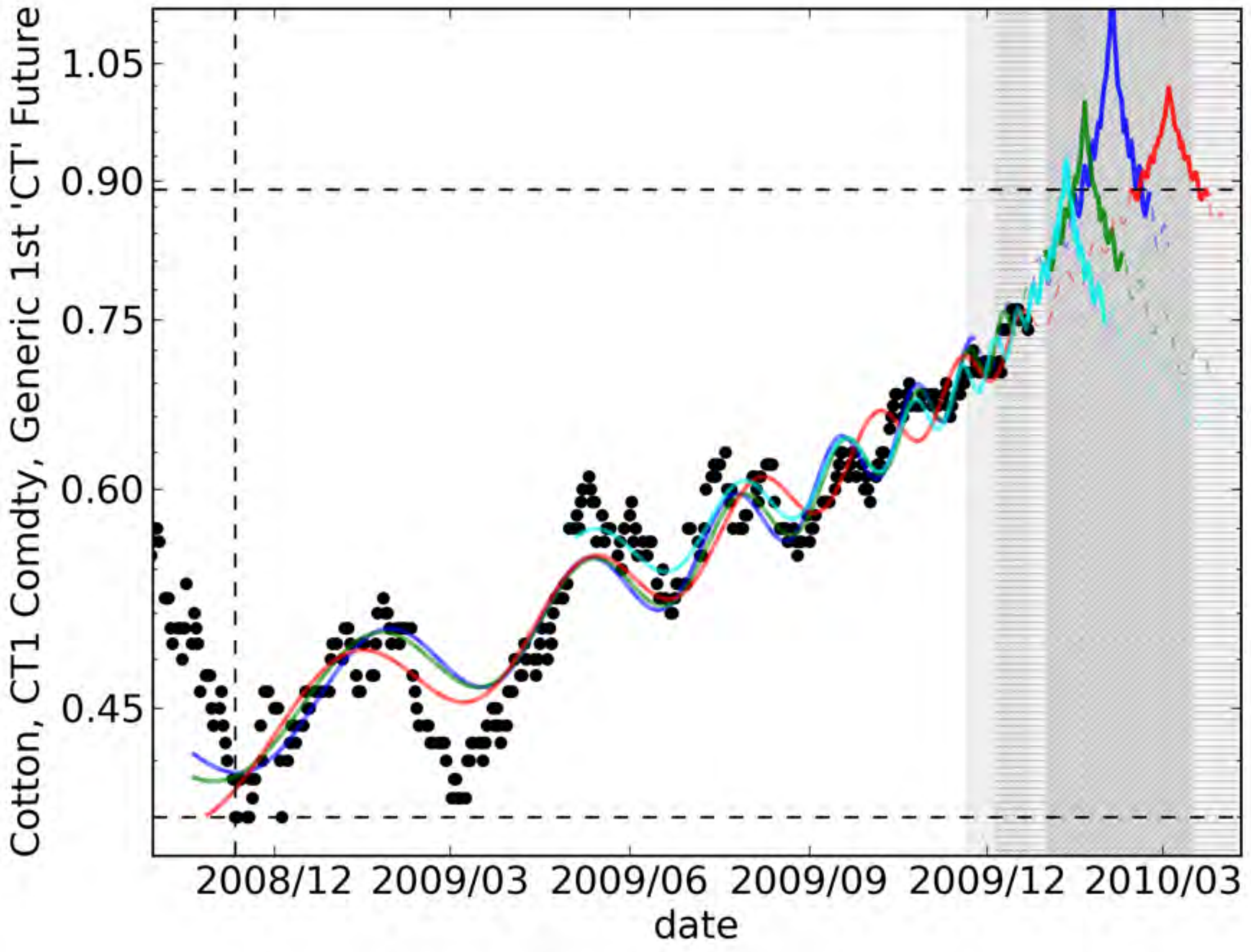

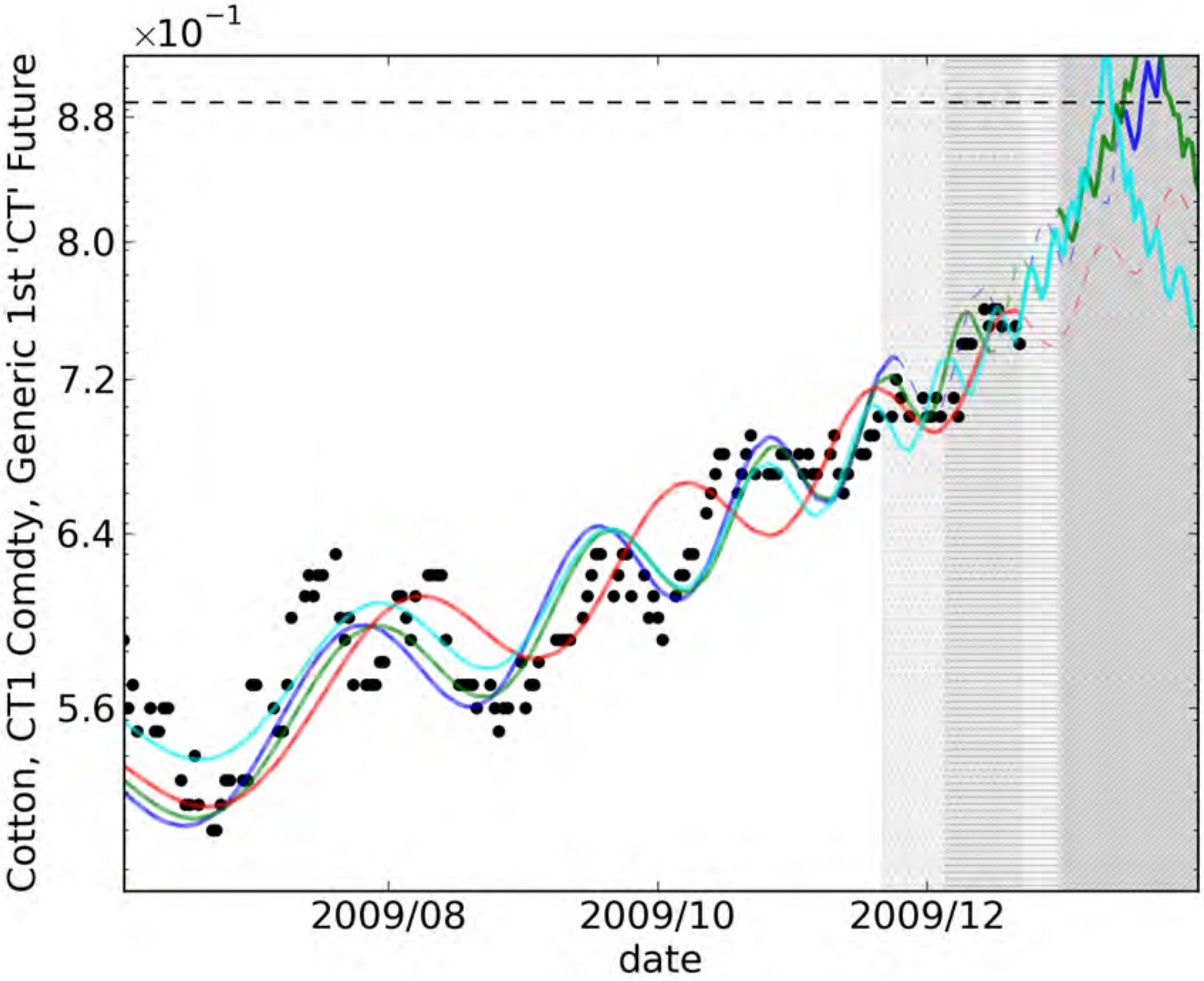

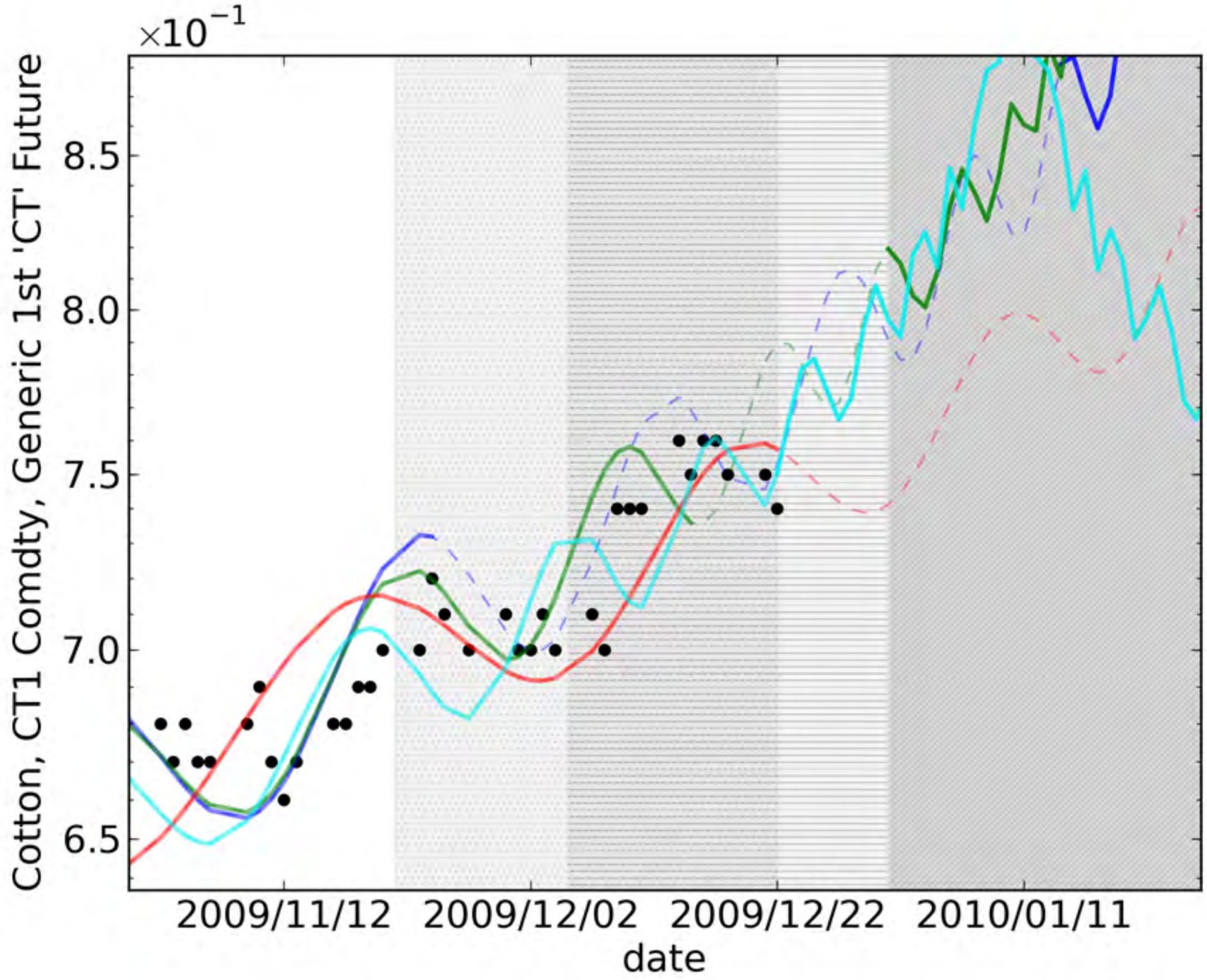